\documentclass[useAMS,usenatbib]{mn2e}

\usepackage[colorlinks]{hyperref}
\usepackage{amssymb}
\usepackage{amsmath}
\usepackage{graphicx}
\usepackage{xspace}
\usepackage{url}
\usepackage{epstopdf}

\usepackage[colorinlistoftodos]{todonotes}

\def\flat{\textit{Fermi}/LAT\xspace}
\def\xrt{\textit{Swift}/XRT\xspace}

\def\cha{\textit{Chandra}\xspace}
\def\nus{\textit{NuSTAR}\xspace}

\def\psrj{PSR~J2032+4127\xspace}
\def\psrb{PSR~B1259-63\xspace}
\def\hessj{HESS~J0632+057\xspace}

\begin{document}
\title[Observations of \psrj during the 2017 periastron passage]{Multi-wavelength observations of \psrj during the 2017 periastron passage}
\author[M. Chernyakova et al.]{M. Chernyakova,$^{1,2}$\thanks{E-mail: masha.chernyakova@dcu.ie} 
          D. Malyshev,$^{3}$ P. Blay,$^{4}$ B. van Soelen,$^{5}$ S. Tsygankov$^{6,7}$
          \\
$^{1}$ School of Physical Sciences and CfAR, Dublin City University, Dublin 9, Ireland\\
              %\email{}\\
$^{2}$ Dublin Institute for Advanced Studies, 31 Fitzwilliam Place, Dublin 2, Ireland\\
$^{3}$ Institut f{\"u}r Astronomie und Astrophysik T{\"u}bingen, Universit{\"a}t T{\"u}bingen, Sand 1, D-72076 T{\"u}bingen, Germany \\
$^{4}$ Valencian International University, Carrer del Pintor Sorolla, 21, 46002 Valencia, Valencia, Spain\\
$^{5}$ University of the Free State, Department of Physics, PO Box 339, 9300 Bloemfontein, South Africa\\
$^{6}$ Department of Physics and Astronomy,  FI-20014 University of Turku, Finland \\
$^7$ Space Research Institute of the Russian Academy of Sciences, Profsoyuznaya Str. 84/32, Moscow 117997, Russia }

\date{Received $<$date$>$  ; in original form  $<$date$>$ }

\label{firstpage}
\pagerange{\pageref{firstpage}--\pageref{lastpage}} 
\pubyear{2019}

\maketitle

\begin{abstract}
    %\psrj is the second known system where a millisecond radio pulsar is  orbiting a massive Be-star. 
    \psrj is only the second known gamma-ray binary where it is confirmed that a young radio pulsar is in orbit around a Be-star.
    The interaction of the pulsar wind with the mass outflow from the companion leads to broad band emission from  radio up  to TeV energies. In the current paper we present results of optical monitoring of the 2017  periastron 
    passage with the Nordic Optical Telescope. These observations are complemented by X-ray (\xrt, \nus) and GeV (\flat) monitoring.  Joint analysis of the evolution of the parameters of the H\,$\alpha$ line and the broadband (X-ray to TeV) spectral shape allows us to propose a model linking the observed emission to the interaction of the pulsar and Be-star winds under the assumption of the inclined disc geometry. Our model allows the observed flux and spectral evolution of the system to be explained in a self-consistent way.
\end{abstract}


\section{Introduction}
\label{sec:intro}

Gamma-ray binaries are a small, but growing class of high mass binary systems, which are characterized by non-thermal emission which peaks in the gamma-ray regime \citep[$>1$\,MeV, see e.g.][]{dubus13}. To date there are eight known systems, %about ten such systems are known,  
all of which consist of an O or B types star and a compact object in the mass range of a neutron star or black hole \citep[e.g.][]{ chernyakova19, corbet19, dubus13}. While the compact object is, for most systems, not directly detected, it is most likely a young, non-accreting radio pulsar and the non-thermal emission is produced in the shock that forms between the pulsar and stellar winds.  This is known to be the case for two systems, namely PSR~B1259-63 and PSR~J2032+4127, where pulsed emission has been detected \citep{johnston92,fermipulsars09}. 

Shortly after the \textit{Fermi} discovery of \psrj as a $\approx143$\,ms gamma-ray pulsar, it was also detected to be a pulsar at radio frequencies  \citep{Camilo09}. The source lies close to the extended TeV HEGRA source, TeV J2032+4130  \citep{hegra05}, and it was proposed that this was a pulsar wind nebula powered by \psrj \citep{aliu14}. However, while \psrj was first thought to be an isolated pulsar \citep{Camilo09}, further radio observations demonstrated a rapid increase in its observed spin-down rate, which was interpreted as evidence that the pulsar is a member of a highly-eccentric binary system, where the optical companion is a $\sim$ 15 M$_{\sun}$ Be star, MT91 213 \citep{lyne15}. \psrj thus turned out to be similar to \psrb, where a 48\,ms pulsar is orbiting around a Be star, LS~2883, in an eccentric, 3.4 years orbit \citep[e.g.][]{johnston92}.

Further multi-wavelength monitoring of \psrj by \citet{ho17} refined the orbital period  to be in the range of $16\,000$ to $17670$ days, with an eccentricity of $e = 0.961$ (separation at periastron is $\alpha_{\rm per}\sim 1$\,au)  and found the  periastron passage would occur in November 2017. Subsequent observations around this period were in  good agreement with this prediction, and in this paper we adopt  $t_{\rm per}$=MJD 58069 as the date of the 2017 periastron.  

\textit{Chandra} X-ray observations reported in \citet{ho17} demonstrated a strong rise in the  source brightness;  \psrj was about 20 times brighter in 2016 than it was in 2010, and about 70 times brighter than it was in 2004. This increase was interpreted as a result of the collision between pulsar and Be star winds. Smoothed-particle hydrodynamics (SPH) modelling of the winds interaction presented in \cite{coe19} demonstrates the deformation and even partial destruction of the Be star disc at some orbital phases.

Modelling of the source suggested that around periastron the binary should be detectable at TeV energies \citep{bednarek18,takata17} and \psrj was subsequently detected at TeV gamma-ray energies by the VERITAS and MAGIC telescopes as a point-like source \citep{psrj_tev}. The TeV light curve peaked at periastron (followed by a short dip a few days later). The observations also showed that during the brightest period (at periastron) the spectrum is best fit by a power-law but when the source was fainter during the 2017 observations (before periastron) a power-law with an exponential cut-off is favoured.

Long term observations of the Be optical companion, MT91 213, have shown the star has demonstrated strong variability showing periods of marked changes in the emission lines. Earlier observations (during 2012 and 2013) showed the stellar spectrum changed between a normal B star to a Be star \citep{salas13} while \citet{ho17} showed that the H\,$\alpha$ equivalent width varied from $W_\lambda = -13.2$\,\AA{} (in 2009) down to $W_\lambda \approx -4$\,\AA{} (in 2014-early 2016). It subsequently showed a rise to   $W_\lambda \approx -10$\,\AA{} before decreasing again over a $\sim200$ day period in the second half of 2016 \citep{ho17,coe19}. Around $\sim200$\,d before periastron, it is seen the equivalent width smoothly decreases towards periastron \citep{coe19,kolka17,roucoescorial19}. 

In this paper we present results of intensive optical observations of \psrj with the Nordic Optical Telescope (NOT) starting approximately 100 days before periastron and lasting up to 50 days after it. These observations were complemented by X-ray (\xrt, \nus) and GeV (\flat) monitoring. Joint analysis of the evolution of the parameters of the H\,$\alpha$ line and the broadband (X-ray to TeV) spectral shape allow us to propose a model linking the observed emission to the interaction of the pulsar and Be-star winds.

%%%%%%%%%%%%%%%%%%%%%%%%%%%%%%%%%%%%%%%%%%%%%%%%%%%%%%%%%%%%%%%%%%%%%%%%%%%%%%%%%%%%%%%%%%%%%%%%%%%%%%%%%%%
\section{Data analysis}
\label{sec:data_analysis}
%%%%%%%%%%%%%%%%%%%%%%%%%%%%%%%%%%%%%%%%%%%%%%%%%%%%%%%%%%%%%%%%%%%%%%%%%%%%%%%%%%%%%%%%%%%%%%%%%%%%%%%%%%%
\subsection{\flat data analysis}
\label{sec:lat_data_analysis}
%%%%%%%%%%%%%%%%%%%%%%%%%%%%%%%%%%%%%%%%%%%%%%%%%%%%%%%%%%%%%%%%%%%%%%%%%%%%%%%%%%%%%%%%%%%%%%%%%%%%%%%%%%%
\flat data selected for the analysis presented in this paper cover more than 11 years (Aug. 2008 to Dec. 2019). We used the latest available \texttt{fermitools}  with P8\_R3 response functions (\texttt{CLEAN} photon class).\footnote{See \href{https://fermi.gsfc.nasa.gov/ssc/data/analysis/documentation/Cicerone/Cicerone_LAT_IRFs/c }{description of \flat response functions} }

%To extract the spectra for all cases below 
To extract the spectra we performed the standard binned likelihood analysis of a region around \psrj. The spectral analysis is based on the fitting of the spatial/spectral model of the sky region around the source of interest to the data. The region-of-interest considered in the analysis is a circle with a radius of 18 degrees around \psrj.  The model of the region included all sources from the 4FGL catalogue~\citep{lat_4fgl} as well as components for isotropic and galactic diffuse emissions given by the standard spatial/spectral templates \texttt{iso\_P8R3\_CLEAN\_V2.txt} and \texttt{gll\_iem\_v07.fits}. 

The spectral template for each 4FGL source in the region was selected according to the catalogue model. The normalizations of the sources were considered to be free parameters during the fitting procedure. Following the recommendation of the \flat collaboration, we performed our analysis with energy dispersion handling enabled. %All upper limits presented in this work were extracted for $TS<4$ cases with the \texttt{UpperLimits} python module provided with the \fermi/LAT software and correspond to the 95 per cent ($\simeq 2\sigma$) false-chance probability.

%%%%%%%%%%%%%%%%%%%%%%%%%%%%%%%%%%%%%%%%%%%%%%%%%%%%%%%%%%%%%%%%%%%%%%%%%%%%%%%%%%%%%%%%%%%%%%%%%%%%%%%%%%%
\subsection{X-ray analysis}
%%%%%%%%%%%%%%%%%%%%%%%%%%%%%%%%%%%%%%%%%%%%%%%%%%%%%%%%%%%%%%%%%%%%%%%%%%%%%%%%%%%%%%%%%%%%%%%%%%%%%%%%%%%
\subsubsection{\xrt}
We accompanied the analysis of the source behaviour with historic \xrt observations of the region and the data from our recent \xrt monitoring campaign of \psrj. The data were reprocessed and analysed as suggested by the \xrt team\footnote{See e.g. the \href{https://swift.gsfc.nasa.gov/analysis/xrt_swguide_v1_2.pdf}{\xrt User's Guide}} with the \texttt{xrtpipeline v.0.13.4} and \texttt{heasoft v.6.22} software package. The spectral analysis of \xrt spectra was performed with \texttt{XSPEC v.12.9.1m}.
%%%%%%%%%%%%%%%%%%%%%%%%%%%%%%%%%%%%%%%%%%%%%%%%%%%%%%%%%%%%%%%%%%%%%%%%%%%%%%%%%%%%%%%%%%%%%%%%%%%%%%%%%%%
\subsubsection{\nus}
Three \nus observations of \psrj were performed on MJDs~57640, 58062 and 58078. The raw data were processed with the standard pipeline processing (HEASOFT v.6.22 with the NuSTAR subpackage v.1.8.0). We applied strict criteria for the exclusion of data taken in the South Atlantic Anomaly (SAA) and in the ``tentacle''-like region of higher background activity near part of the SAA. Level-two data products were produced with the \texttt{nupipeline} tool with the flags \texttt{SAAMODE=STRICT} and \texttt{TENTACLE=yes}. High-level spectral products (spectra, response matrices, and auxiliary response files) were extracted for a point source with the \texttt{nuproducts} routine. The corresponding background flux was derived from a ring-like (inner/outer radii of 39''/86'') region surrounding the source. The spectral analysis was performed in the energy range of $3-70$~keV.

%%%%%%%%%%%%%%%%%%%%%%%%%%%%%%%%%%%%%%%%%%%%%%%%%%%%%%%%%%%%%%%%%%%%%%%%%%%%%%%%%%%%%%%%%%%%%%%%%%%%%%%%%%%
\subsection{Optical spectroscopy}
\begin{figure}
    \centering
    \includegraphics[width=0.99\columnwidth]{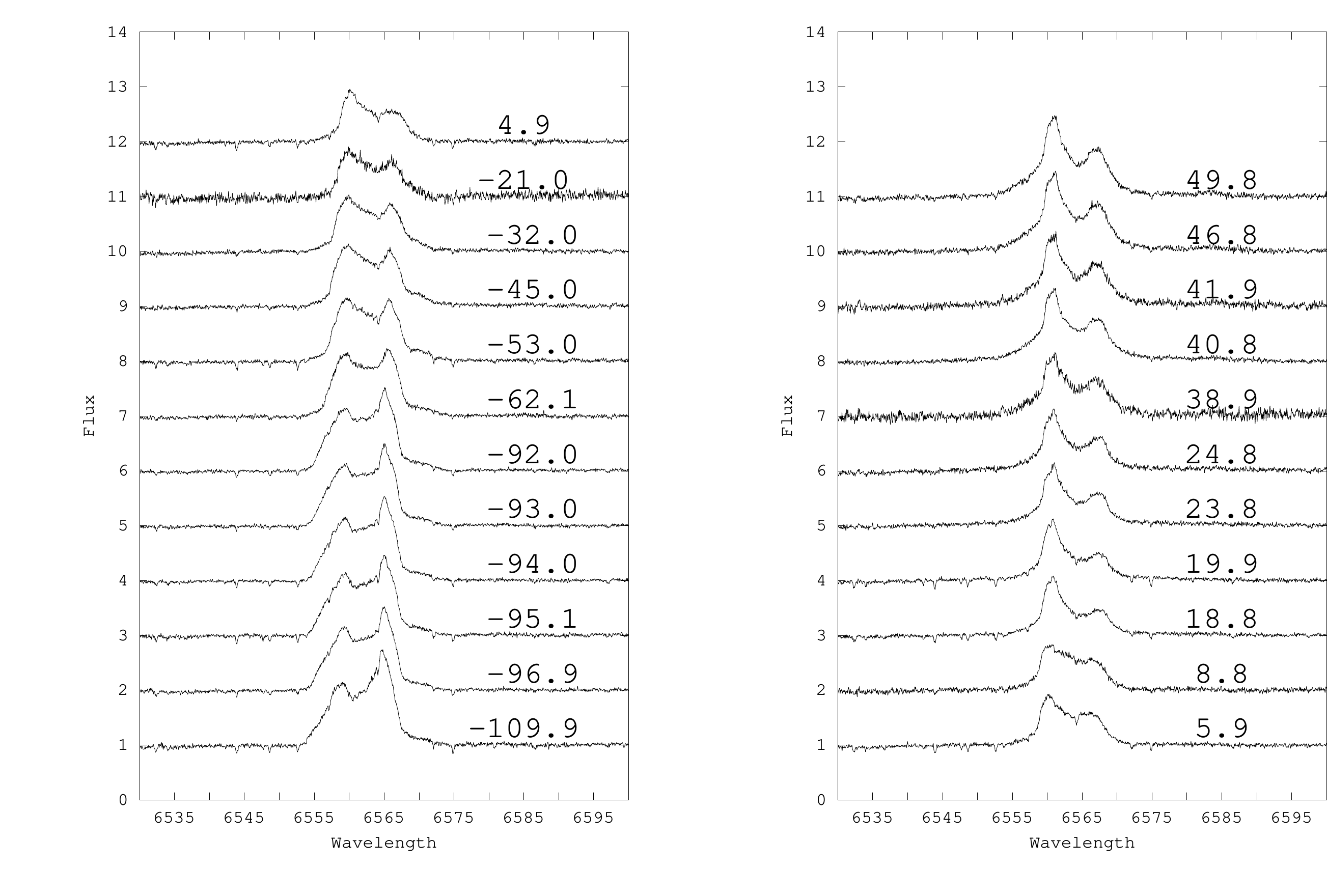}
    \caption{Evolution of the H\,$\alpha$ line profile. Time runs upwards and from left to right. The difference (in days) between MJD and periastron passage (58069) is shown on the right side of each profile. Details for each spectrum are shown in Table \ref{tab:opt1}.}
    \label{fig:opt1}
\end{figure}

%\todo[inline]{Describe optical measurements/data analysis, etc}
High resolution (R=25000) spectra were taken with the NOT from MJD 57950 to 58120, covering almost the whole periastron passage of the target. The instrument used was the FIber-fed Echelle Spectrograph (FIES, Telting et al. 2014) in its lower resolution mode. The spectra cover the range from 3800 up to 8300~\AA. Exposure times were optimised for a good signal to noise ratio in the H\,$\alpha$ line region, therefore the signal to noise in the blue bands is very low. This, together with the emission lines filling in all visible Balmer lines in that range, made the blue part of the spectra unusable for spectral characterisation of the target. A summary of the spectra and signal to noise ratio around the H\,$\alpha$ line is shown in Table \ref{tab:opt1}. The history of H\,$\alpha$ profiles is shown in Fig.~\ref{fig:opt1}.

\begin{table}
    \centering
        \caption{Summary of H\,$\alpha$ parameters (namely, equivalent width in \AA, peak ratio and peak distance in km\,s$^{-1}$) are shown, together with the middle MJD time relative to periastron,  signal to noise (S/N) ratio and exposure time (EXP.). The given S/N  refers to the continuum around the H\,$\alpha$ line. }
    \label{tab:opt1}
    \begin{tabular}{@{}c@{}|c@{}|c@{}|c@{}|c@{}|c@{}|c@{}} \hline
    MJD & MJD-$T_{per}$ & S/N & EXP. & EW  & V/R & R-V  \\
        &             &     &   (ks) & (\AA)&   & (km~s$^{-1}$) \\
    \hline
57959.15 & -109.9 & 50 & 2.5 & -12.91$\pm$0.12 & 0.66$\pm$0.01 & 277$\pm$3	\\
57972.14 & -96.0 & 50  & 2.7 & -12.20$\pm$0.05 & 0.78$\pm$0.03 & 262$\pm$2      \\
57973.95 & -95.1 & 50  & 2.7 & -11.51$\pm$0.10 & 0.76$\pm$0.03 & 259$\pm$3      \\
57975.00 & -94.0 & 50 & 2.7 & -12.10$\pm$0.10 & 0.75$\pm$0.02 & 264$\pm$2      \\
57976.05 & -93.0 & 50 & 2.7 & -11.85$\pm$0.11 & 0.81$\pm$0.05 & 268$\pm$4      \\
57977.05 & -92.0 & 50 & 2.7 & -12.16$\pm$0.11 & 0.74$\pm$0.03 & 256$\pm$3      \\
58006.90 & -62.1 & 60 & 1.2 & -11.19$\pm$0.06 & 0.95$\pm$0.05 & 287$\pm$3      \\
58016.04 & -53.0 & 45 & 2.5 & -10.68$\pm$0.08 & 1.02$\pm$0.05 & 292$\pm$3      \\
58023.98 & -45.0 & 50 & 1.2 & -10.08$\pm$0.06 & 1.11$\pm$0.02 & 297$\pm$1      \\
58036.99 & -32.0 & 50 & 1.2 &  -8.47$\pm$0.06 & 1.12$\pm$0.03 & 298$\pm$4      \\
58048.00 & -21.0 & 25 & 1.2 &  -6.80$\pm$0.11 & 1.32$\pm$0.13 & 297$\pm$3      \\
58073.89 & 4.9 & 55 & 1.2  & -6.46$\pm$0.03 & 1.90$\pm$0.09 & 343$\pm$2      \\
58074.91 & 5.9 & 55 & 1.2  & -6.44$\pm$0.06 & 1.67$\pm$0.14 & 319.8$\pm$0.4	\\
58077.83 & 8.8 & 35 & 1.2  & -6.22$\pm$0.04 & 1.32$\pm$0.10 & 300$\pm$1	   \\
58087.81 & 18.8 & 65 & 1.2  & -6.21$\pm$0.07 & 2.48$\pm$0.27 & 327.5$\pm$0.5	\\
58088.93 & 19.9 & 55 & 1.2  & -6.54$\pm$0.06 & 2.40$\pm$0.09 & 338$\pm$1      \\
58092.84 & 23.8 & 60 & 1.2  & -7.02$\pm$0.05 & 1.92$\pm$0.27 & 330$\pm$1      \\
58093.80 & 24.8 & 40 & 1.2  & -7.04$\pm$0.07 & 1.72$\pm$0.11 & 324$\pm$1      \\
58107.85 & 38.9 & 20 & 1.2  & -8.26$\pm$0.05 & 1.71$\pm$0.14 & 290$\pm$4      \\
58109.81 & 40.8 & 50 & 1.2  & -9.52$\pm$0.06 & 1.67$\pm$0.05 & 305$\pm$1      \\
58110.89 & 41.9 & 30 & 1.2  & -9.13$\pm$0.11 & 1.74$\pm$0.16 & 312$\pm$3      \\
58115.82 & 46.8 & 35 & 1.2  & -10.14$\pm$0.05 & 1.61$\pm$0.13 & 297$\pm$1      \\
58118.81 & 49.8 & 40 & 1.2  & -10.45$\pm$0.08 & 1.67$\pm$0.03 & 292$\pm$1      \\ \hline
    \end{tabular}

\end{table}

In our observations the H\,$\alpha$ line profile is always double-peaked. The reddest peak is more prominent in our first spectrum, but that situation reverses in our last spectrum, where the bluest peak is the dominant one. This transition from one dominant peak to another is typical in Be star discs and is the observational footprint of a density wave within the disc \citep[see e.g.][]{telting94}. 

\begin{figure}
    \centering
    \includegraphics[width=0.98\linewidth]{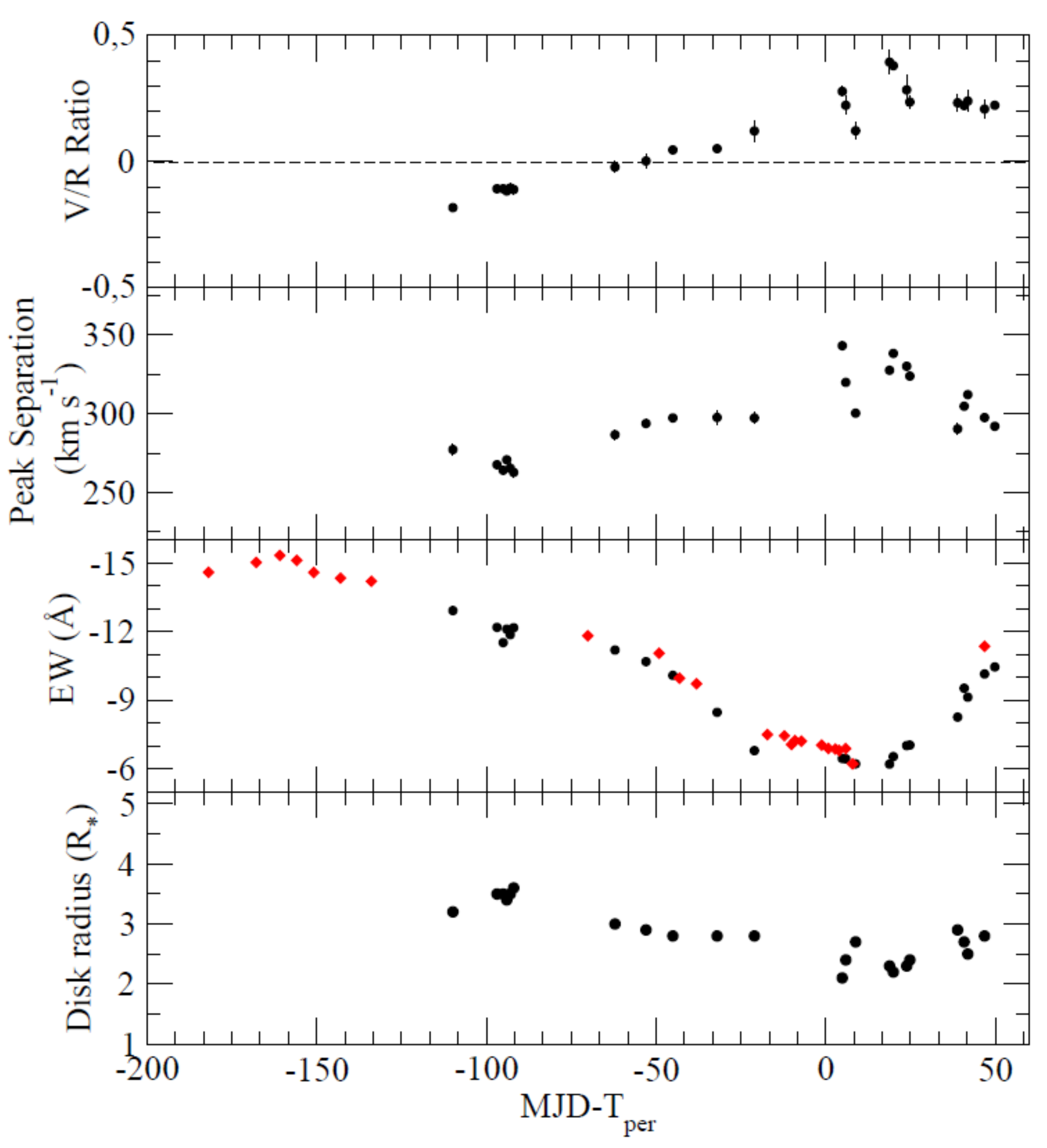}
    \caption{Evolution of the H\,$\alpha$ parameters from the FIES spectra of MT91 213. Equivalent width measurements from \citet{coe19} are shown in red (without errors) for comparison;  the V/R ratio and peak separation are excluded because of the large errors. The lower panel shows the radius of the H\,$\alpha$ emitting region of the disc.}
    %\todo[inline]{Does it have sense to show schematically other(not our) optical observations, to illustrate, that e.g. EW has "double-peak" structure, not just fall and rise?}}
    \label{fig:opt2}
\end{figure}

We have performed a set of measurements of the  H\,$\alpha$ line which include the equivalent width (EW), the peak separation R-V, where R represents the red-most peak position and V the blue-most peak, and the ratio of peak fluxes V/R. In order to calculate the R-V distance in km\,s$^{-1}$, we have used the relationship $(R-V)=c(\lambda_R - \lambda_V)/\lambda_\alpha$, where $c$ is the speed of light and $\lambda_\alpha$ is the rest wavelength of the $H_\alpha$ line.  These parameters characterise the properties of the circumstellar discs around Be stars. The evolution in time of these three parameters can be seen in Fig.~\ref{fig:opt2}. Peak positions and fluxes have been obtained directly from the observed profile, by smoothing and finding each maximum when the slope of the line profile becomes null. All our measurements are in good agreement with those of \cite{coe19}.

We see a decreasing EW as the neutron star approached its periastron. Right after the periastron passage, the disc becomes unstable and all three parameters show large variability. They recover a more steady trend afterwards, with the EW rising towards previous values and both the peak separation and ratio decreasing towards values similar to those before periastron passage. The high degree of variability after the periastron passage may be directly linked to the instabilities created by tidal interactions with the neutron star, as in previous epochs the reported variability is smoother and takes place over longer time scales \citep[see][]{coe19}.

\begin{figure}
    \centering
    \includegraphics[width=0.7\linewidth]{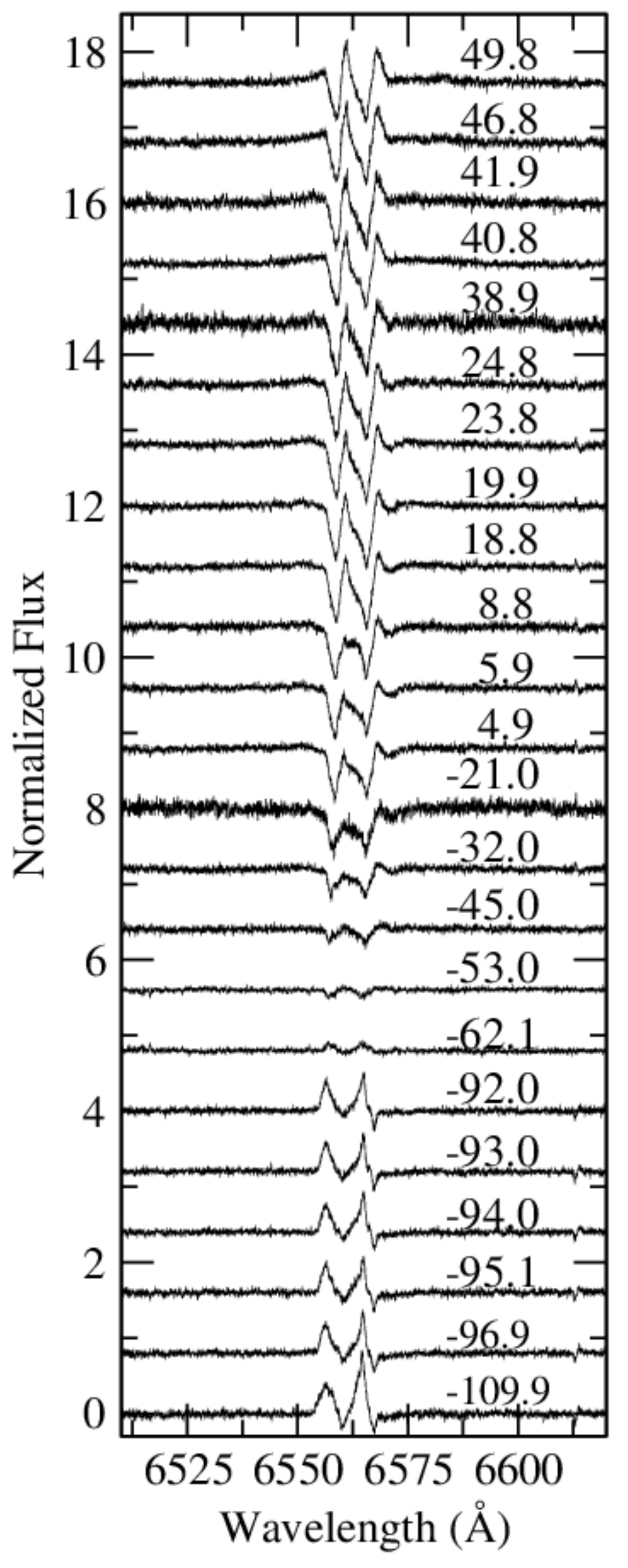}
    \caption{Difference between all spectra in Fig. \ref{fig:opt2} and the average of the two most symmetric ones, i.e., those for which V/R ratio is $\sim$1. We see the emission excess on the wings of the line due to line asymmetry, initially on the blue side but decreasing in intensity as the V/R ration increases. This  excess appears on the red side when V/R $>$ 1 and increases in intensity afterwards. The line wing at the side of the dominant peak falls more abruptly to the continuum than the wing at the opposite side. The excess due to zones of enhanced density propagating on the disc (the cause of the V/R variability), seen in the central zone of these profiles, moves from red to blue.}
    \label{fig:opt3}
\end{figure}

In order to look deeper into the detailed H\,$\alpha$  line structure, we have selected the most symmetric profiles (those at MJD$\sim$58006 and MJD$\sim$58016, when the V/R ratio is $\sim$1), and after averaging them we have subtracted that average from the rest of spectra. The results are shown in Fig.~\ref{fig:opt3}. There we can see the excess in emission moving from the right (red side) towards the left (blue side). We also see some residuals at the line wings which evolve in the opposite direction. These residuals at the wings are a consequence of the H\,$\alpha$ line asymmetry.  This asymmetry is expected for density waves perturbing the disc \citep{hummel97} and are dependent on the inclination of the disc with respect to the line of sight. The asymmetric profile in H\,$\alpha$, according to \citet{hummel97}, would correspond to a disc inclination angle of between 25 and 50 degrees.
% would correspond to an inclination of the disc with respect to the line of sight of between 25 and 50 degrees.

 The peak of the excess in emission responsible of the V/R variability (Fig.~\ref{fig:opt3})
%The peak  of the emission responsible of the V/R variability 
moves from a redder ($t_{\rm per} \sim -109.9$\,d) to a bluer ($t_{\rm per} \sim 49.8$\,d)  position by a distance on the order of $\sim$170\,km\,s$^{-1}$. The peak separation shown in Fig.~\ref{fig:opt2}, with an average of $\sim$290\,km\,s$^{-1}$, is also influenced by the emission excess in the line wings. The bluest peak of this wing excess at  $t_{\rm per} \sim -109.9$\,d differs in position with respect to the reddest one at $t_{\rm per} \sim 49.8$\,d on the order of $\sim$560\,km\,s$^{-1}$. This distance is in agreement with what is expected from the model of \cite{hummel97}. 

The lower panel on Fig. \ref{fig:opt2} shows the evolution of the circumstellar disc radius. In order to estimate the disc radius, first  we have estimated the projected rotational velocity of MT91 213 from the peak separations  (R-V) seen in Table \ref{tab:opt1} by using the relationship
% \begin{center}
% $\Delta {\rm peak} \sim (1.2\pm0.6)v_{\rm rot}\sin i$ 
% \end{center}
\begin{equation}
    \Delta {\rm peak} \sim (1.2\pm0.6)v_{\rm rot}\sin i
\end{equation}
which is valid for H\,$\alpha$ EWs lower than 15 \AA  \citep{hanuschik89}. We obtain a projected rotational velocity of $250\pm20$\,km\,s$^{-1}$. With this projected rotational velocity, and using the relationship \citep[see, e.g., ][]{reig16}
% \begin{center}
%     $\frac{R_{\rm disc}}{R_{*}}=(\frac{2v_{\rm rot}\sin i}{\Delta {\rm peak}})^{1/j}$ 
% \end{center}
\begin{equation}
    \frac{R_{\rm disc}}{R_{*}}=\left(\frac{2v_{\rm rot}\sin i}{\Delta {\rm peak}}\right)^{1/j}
\end{equation}
we have calculated $R_{\rm disc}/R_{*}$. The $j$ in the exponent is expected to be $\sim$0.5 for a Keplerian disc \citep{reig16}. By fitting
% \begin{center}
%     $\log(\frac{\Delta {\rm peak}}{v_{\rm rot}\sin i})=-(\frac{j}{2})\log(-{\rm EW}(H\alpha))+b$
% \end{center} 
\begin{equation}
    \log\left(\frac{\Delta {\rm peak}}{v_{\rm rot}\sin i}\right)=-\left(\frac{j}{2}\right)\log(-{\rm EW}_{{\rm H}\alpha})+b
\end{equation}
\citep[see][]{reig16} to the measurements from all our spectra we obtained $j\sim0.53$. 

\section{Results and discussion}
\subsection{Optical evidence of the Be-star disc evolution.}
%\todo[inline]{I'm trying to merge parts of 3.4/3.5 because of repetition.  }

%%%%%%%%%%%%%%%%%%%%%%%%%%%
%%%%%%%%%%%%%%%%%%%%%%%%%%%

%Despite the fact that the geometry of \psrj, where a pulsar passes through an ``inclined disc'', would suggest that it should behave similarly to several other gamma-ray binaries (e.g. \psrb or \hessj), its behaviour during the periastron passage does not exactly resembles those of the other systems. 

The optical behaviour of \psrj was previously discussed by e.g.\ \citet{coe19, roucoescorial19, kolka17}.  However the majority of these observations occurred before periastron and the cadence of observations after periastron was much less. The NOT observations presented in the paper, complemented by the previous observations, show the behaviour of the disc until $\sim50$\,d after periastron.

The observations of the H\,$\alpha$ line in \psrj show a clear variability of the Be star disc, both in the change in equivalent width, and the V/R profile.  The equivalent width well away from periastron is $|W_\lambda|<9$\,\AA{} \citep{coe19} but becomes stronger $\sim181$\,d before periastron ($|W_\lambda|\sim15$\,\AA{}). The EW then decreases towards periastron reaching a minimum ( $|W_\lambda|\approx 6.2$\,\AA{}) a few days after periastron, while the V/R profile shows a corresponding shift to a stronger blue component and an increase in the peak separation (Fig.\ref{fig:opt1}).

The behaviour observed immediately around and after periastron is easily explained by the introduction of tidal interaction with the pulsar \citep[see the SPH modelling in ][]{coe19}.  However, the behaviour of the disc before and leading up to periastron is more challenging. 

Be stars are known to show intrinsic variability of the disc structure \citep[see e.g.][and references therein]{rivinius13}, and a number of the observed changes in the disc, particularly seen far from periastron, are due to the intrinsic variability of the Be star and not the interaction with the pulsar. However, the clear variation of the EW towards periastron, with the line strength decreasing towards, and increasing after,  periastron is very suggestive that this can be attributed to the interaction in the binary system. 
 It should be noted that \citet{kolka17} suggested that the behaviour of the disc around periastron was due to intrinsic variation of the star and that size of the disc remained stable as they did not measure a shift in the position of the peaks of the H\,alpha line. However, the longer period and higher resolution observations presented here and in \citet{coe19} do find a shift.

% The maximum in the EW occurs $\sim161$\,d before periastron \citep{coe19} following by a decrease  towards periastron. 

% The peak separation shows a decrease towards $\sim100$\,d before periastron, and a following rise towards periastron (see Fig.~\ref{fig:opt2}). 

While the optical emission occurs close to the Be star (see Fig.~\ref{fig:opt2}) 
% \citep[the H\,$\alpha$ line generating region is at size of $\sim 14 R_*$;][]{coe19} 
the disc can be much larger. For example, \citet{klement17} suggested truncation radii of $\sim100$\,$R_*$, for some Be stars, which they attributed to undetected binary companions. In the viscous decretion disc model the density of Be disc decrease with radial distance $\rho \propto r^{-n}$, and this rate is lower of Be stars in binaries due to the accumulation of material within the truncation radius, resulting in denser discs \citep[e.g.][]{zamanov01, okazaki02, panoglou16}. 
%
%While the binary separation is very still large at $\tau \sim -160$\,d ($\approx 7$\,au $\approx 150 R_{*}$ assuming $M_*=10M_\odot$ and $R_*=10 R_\odot$) the highly eccentric nature of this binary may mean that truncation causes a slight accumulation in material closer to the star, increasing the EW and moves the dominant emission region slightly further out. 
%
The highly eccentric nature of this binary may mean that as the pulsar approaches the star truncation of the disc slows the rate at which material can flow out of the decretion disc,  causing an accumulation in material close to the star, increasing the EW. This may explain the rise in equivalent width from $|W_\lambda|<9$\,\AA{} to $|W_\lambda|\sim15$\,\AA{} between the observations before $t_{\rm per}-393$\,d and the next after $\sim t_{\rm per} - 181$\,d shown in \citet{coe19} (where $t_{\rm per}$ is the time of periastron).   There is also a clear change in the line profile with red peak significantly increasing \citep[see fig.~5 in][]{coe19}. The gravity of the pulsar could further be causing a preferential direction for the disc outflow. However, it should be noted that at these times the binary separation is still large and underlying intrinsic variability of the Be star be driving this increase \citep{kolka17}. 

The observations after $\sim t_{\rm per} -181$\,d \citep[in ][and this paper; see Fig.~\ref{fig:opt2}]{coe19} show that the equivalent width peaks at $\sim t_{\rm per}-161$\,d and then steadily decreases towards periastron. The NOT observations (starting from $\sim t_{\rm per}-110$\,d) also show a general increase in the peak separation. Since Be stars are believed to be Keplerian \citep[e.g.][]{rivinius13} the peak separation traces the %dominant 
dominant or average H\,$\alpha$ emission region, and the change suggests the emission region is moving closer to the Be star. % as the pulsar approaches periastron. 
This could be occurring because as the truncation of the disc increases, the emitting region becomes smaller, decreasing the EW and moving the average emission region slightly closer to star. The disc truncation could be induced by tidal effects, but additionally the pulsar wind may be blowing away regions of the disc. This  behaviour is very different to the case of \psrb\ where the equivalent width continues to grow past periastron.  However, the typical equivalent width of the H\,$\alpha$ line observed from \psrb\ is much higher ($|W_\lambda|\sim 50$\,\AA{}) which suggests a much denser disc, than in the case of \psrj.

% A less dense disc would be more easily removed from around the star. 

%We can also estimate the radius of where the emission occurs by...
%\todo[inline]{Estimate the change in radius. Need typical rotation speed of star. Could we use the breakup velocity?}

%The radius of the emitting region is estimated using \citep{huang72}
%\[
%\frac{R}{R_\star}  = \left( \frac{2 v \sin i}{\Delta v} %\right )^2
%\]
%where we have assumed the star is rotating at 90 per cent of %the critical speed given by
%\[
%v_{\rm crit} = \sqrt{\frac{2}{3} \frac{G M}{R}  }
%\]
%for $M = 10 M_\odot$ and $R = 10 R_\odot$ 

% %%%%%%%%
% \begin{figure}
%     \centering
%     \includegraphics[width=\linewidth]{comparison.pdf}
%     \caption{Temporary figure to compare new results to Coe+19}
%     \label{fig:comparison}
% \end{figure}

%Further, the V/R profile of the H\,$\alpha$ lines shows a line shifts from being symmetric at $t_{\rm per}=-600$ to $-400$\,d \citep{coe19}, to varying from redder to more violet over the periastron passage. 
The influence of the pulsar is also seen in the change in the V/R profile.  While the double peaked line was symmetric ($V/R \approx 1$) far from periastron \citep[ $t_{\rm per}-600$ to $t_{\rm per}-400$\,d;][]{coe19}, after $\sim t_{\rm per}-182$\,d  the line is asymmetric and varying from redder to bluer. 
The variation in V/R profile is normally attributed to density waves in the decretion disc, which in this case can be induced by the tidal interaction as is seen in SPH modelling of Be binary system \citep{okazaki02,panoglou16,coe19}.
The variation in the V/R ratio changes relatively smoothly as the pulsar approaches periastron.
%As the pulsar moves towards and past periastron, the variation in the disc changes, with the V/R ratio shifting from a stronger red to a stronger blue component as well as showing more rapid variability in the disc V/R profile.
However, immediately after periastron (within a few ten of days) the variation in the V/R profile and peak separation changes much more rapidly (Fig.~\ref{fig:opt2}). The peak separation reaches a maximum of 343\,km\,s$^{-1}$ before decreasing by $\sim 40$\,km\,s$^{-1}$  in $\sim3$\,d, before increase by  $\sim 40$\,km\,s$^{-1}$  within $\sim2$\,d, before decreasing again.  At the same time the V/R profile first becomes slightly redder before becoming bluer again. Because the system is highly eccentric, the binary separation at periastron is $\sim1$\,au ($\sim 28\,R_\star$), and the pulsar will introduce a strong turbulence to the disc as well as an asymmetry to the disc structure.  As the pulsar passes periastron more turbulent structures will be introduced into the disc and material can be pulled in the direction of the pulsar, then continuing to move towards us, resulting in a stronger blue component. The disc will become less symmetric and change more rapidly, causing the faster change in the V/R ratio and peak separation. This basic structure is seen in the  SPH models presented in \citet{coe19} where the disc becomes asymmetric and material is carried away from the disc.  

During the post-periastron period the equivalent width starts to increase and we interpret this as the disc starting to increase in size as the binary separation increases.  

\subsection{High Energy Emission}

%%%%%%%%%%%%%%%%%%%%%%%%%%%%%%%%%%%%%%
\begin{figure*}
\includegraphics[width=0.48\linewidth]{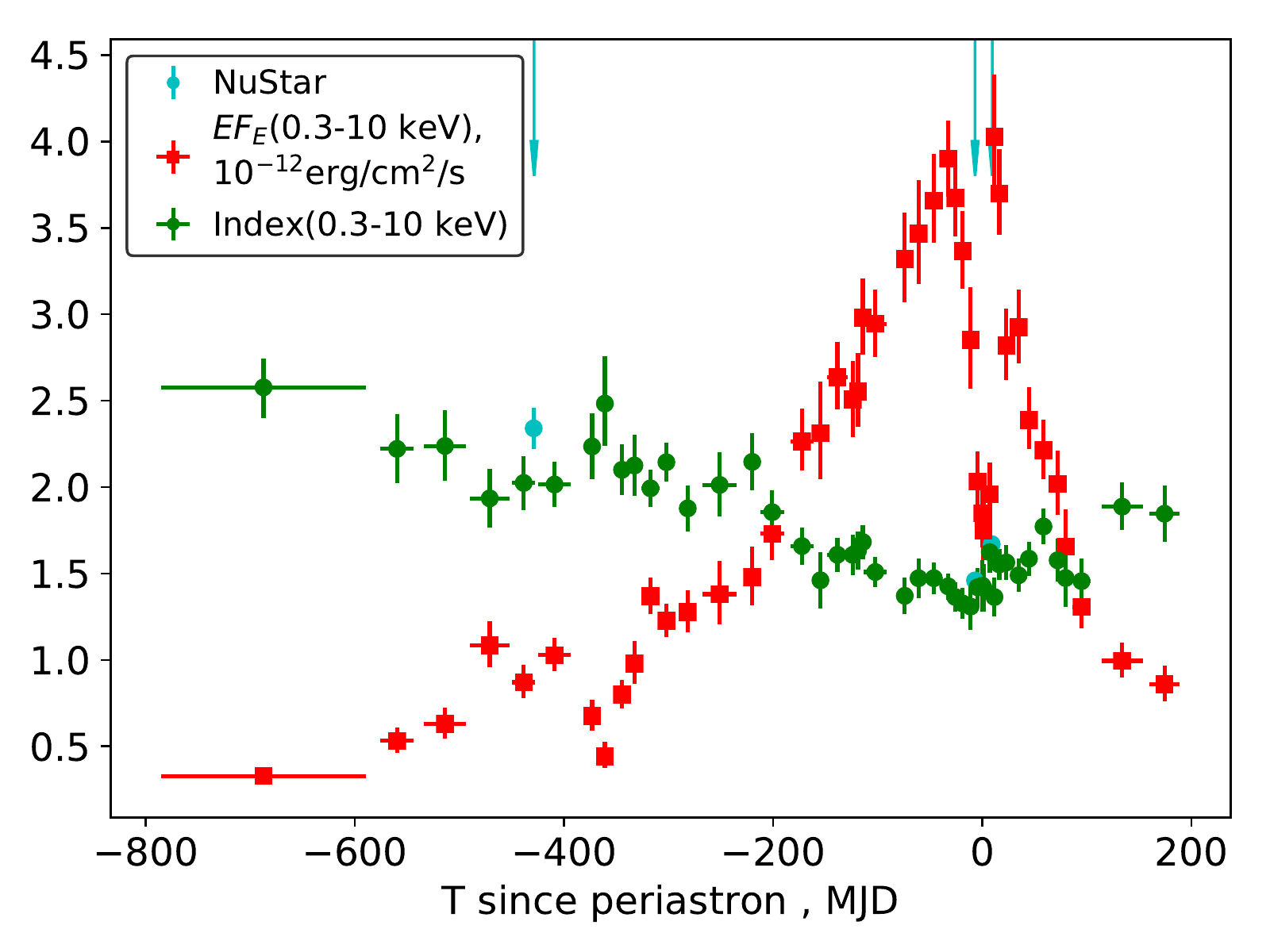}
\includegraphics[width=0.48\linewidth]{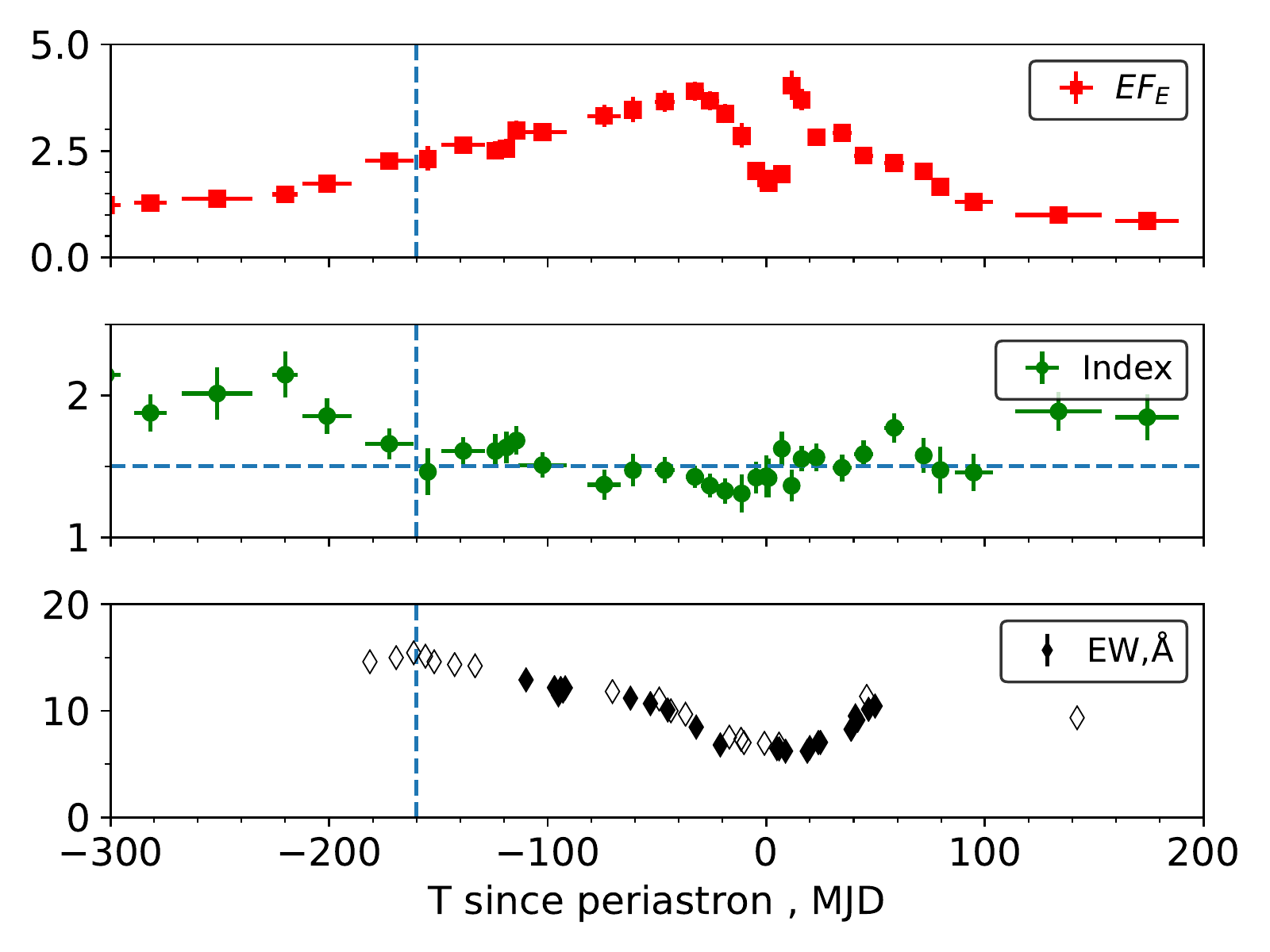}
\caption{  \xrt 0.3-10~keV lightcurve of \psrj (red squares) and the corresponding slope (green circles). Cyan points and arrows demonstrate index ($3-70$~keV) and periods of the {\it NuSTAR} observations.  Black symbols in the right panel illustrate the equivalent width of the H\,$\alpha$ line reported in this paper (filled symbols) and~\citet{coe19} (empty symbols). Vertical dashed line denotes the period of the peak of H$\alpha$ equivalent width. Horizontal  dashed line corresponds to the value of the X-ray slope equal to 1.5.}
\label{fig:xrt_lc_index}
\end{figure*}
%%%%%%%%%%%%%%%%%%%%%%%%%%%%%%%%%%%%%% 
While the optical emission detected from \psrj is mainly coming from the Be star and its disc, the observed X-ray emission is usually attributed to the synchrotron emission of the relativistic electrons originating from the %of the 
pulsar wind. The first X-ray detection of this system was in 2002 by 
\cha \citep{Camilo09,ho17}, which reported the detection of a new, very faint X-ray source at the position of  \psrj. Subsequent observations with %Chandra and SWIFT 
\cha and \xrt demonstrated a steady flux rise as the pulsar approached periastron.   
The dependencies of the X-ray flux and slope on time are summarized in Fig.~\ref{fig:xrt_lc_index}.
The flux demonstrates a complicate behaviour with a sudden narrow dip close to periastron. The slope of the X-ray band spectrum hardens as the pulsar approaches the Be star, reaching a value of $\Gamma\sim1.5$ and remains constant around the periastron passage, see also \citet{Pal19} for the discussion of the spectral slope behaviour around the periastron. %(true anomaly $\phi=-130..130$). 
Note that no significant variation of the hydrogen column density was detected and for the presented analysis it was fixed to its mean value over the orbit,  $N_H=0.884\cdot 10^{22}$~cm$^{-2}$. The correctness of the X-ray slopes measured by \xrt  (which can be biased due to a correlation with $N_H$) is supported by \nus measurements. These measurements performed at $>3$~keV energies are minimally affected by the absorption and are in a good agreement with the \xrt results.

\begin{figure}
\includegraphics[width=\linewidth]{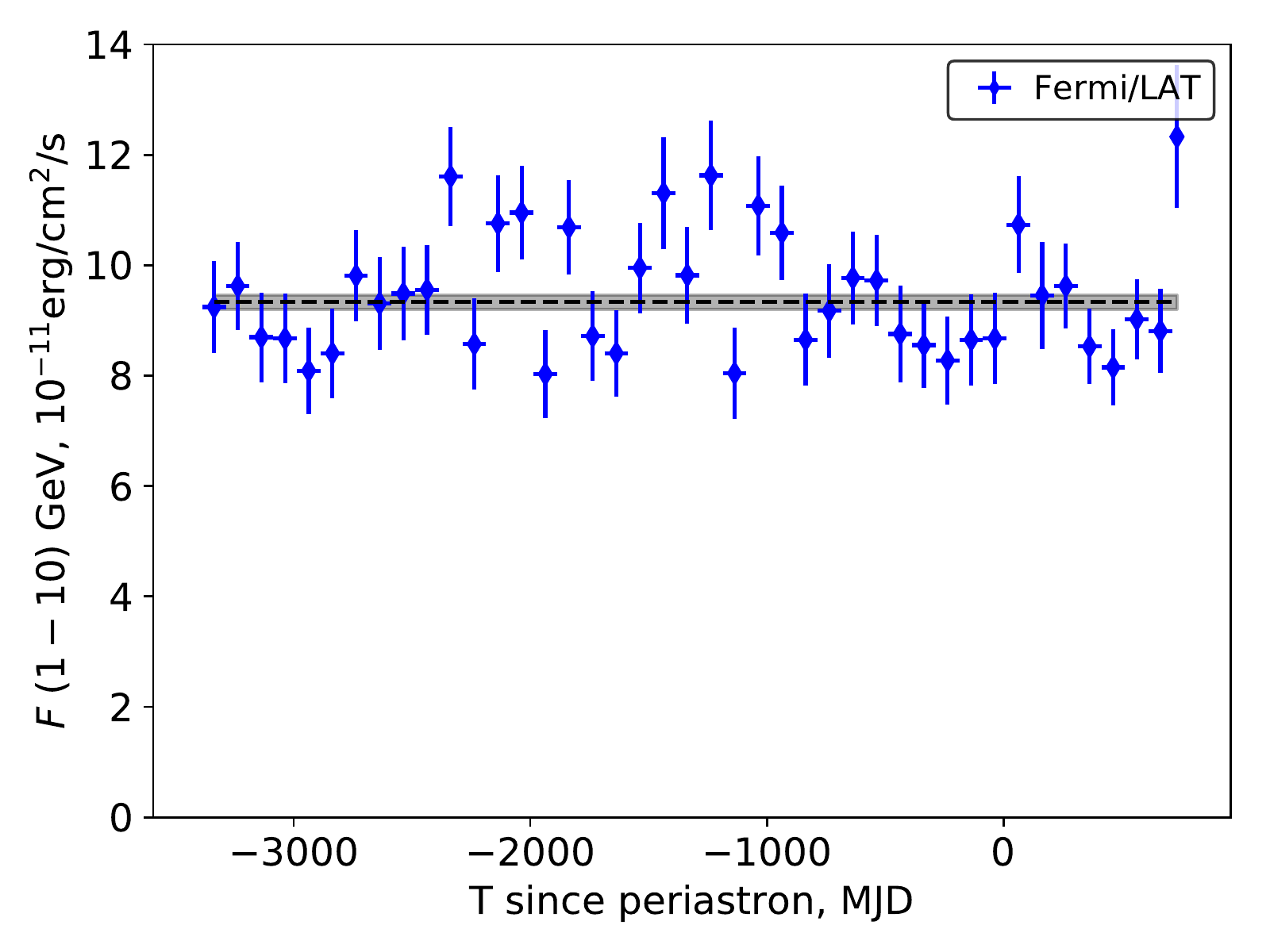}
\caption{ \flat lightcurve of \psrj (1-10 GeV) with a 100~days bins. The black dashed line and shaded region illustrate the best-fit of the data with a constant and the corresponding $1\sigma$ uncertainty area. }
\label{fig:lat_lc}
\end{figure}

Contrary to the X-ray band, the flux and the spectral slope of \psrj remained stable in the GeV band for the whole observational period including the periastron passage, see  \citet{li18} and Fig.~\ref{fig:lat_lc}.

Previously, the time evolution of \psrj was modelled by~\citet{li18} (radio, X-ray and GeV bands) and~\citet{coe19}(optics and X-ray).  
\citet{li18} suggested that the strong X-ray dip close to periastron is explained by an increase of the magnetization parameter of the star--pulsar colliding winds shock accompanied by flux suppression due to Doppler boosting effect. The %following 
post-periastron rise could be a consequence of the Be stellar disc passage by the pulsar. The absence of variability in the GeV emission was explained by the strong dominance of the pulsar magnetospheric emission over the expected orbital-modulated Inverse Compton (IC) emission. The model was able to predict the overall shape of the orbital X-ray lightcurve, but was not able to reproduce the details of the double-peak flux structure around periastron. The origin of the hardening of the X-ray spectrum, as the pulsar approaches periastron, also does not appear in the model in a natural way and was attributed by the authors to a possible increase of the hydrogen column density.\footnote{Please note, that only the hardness ratio was evaluated by~\citet{li18} .}
%  $N_H$.

\citet{coe19} reported on optical and X-ray flux measurements of \psrj,  accompanied by  SPH modelling of the Be star/pulsar interaction. In their model the authors explicitly assumed that the disc of the Be star is inclined to the orbital plane. The modelling, however, failed to describe the details of the observed X-ray lightcurve of the system, generally predicting a maximum of the flux at periastron and % lacks the spectral variability studies.
does not consider the variability of the X-ray spectrum. 

Additionally, \citet{ng19} recently presented radio to X-ray observations of \psrj. The authors suggested a potential spectral break at $\sim 5$~keV energies which was attributed to the modification of the spectrum of accelerated electrons by synchrotron losses without detailed modelling. % of the processes in the system. 

%\todo[inline]{Discuss radio/X-ray observations of \citet{ng19}. Say that unpulsed radio is %stable along the orbit. Pulsed -- demonstrates clear variability. Did they model something?}

Below we propose a simple model for the observed emission from \psrj  which allows  the observed multi-wavelength evolution %of the properties 
of this system to be  qualitatively explain. Within this model, similar to~\citet{coe19}, we suggest that the disc of the Be star is inclined to the orbital plane. 

We would like to note that the inclination of the disc to the orbital plane was also proposed for other gamma-ray binary systems, e.g. \psrb~\citep{melatos95,chernyakova06} and HESS J0632+057~\citep{we_hessj}. These systems demonstrate similar lightcurve with two peaks in X-ray band separated by a dip at periastron and  are characterised by a long (longer than a year) orbital period. 
Such similarities allow us to speculate, that the ``inclined disc'' geometry can be typical for long-period Be-star hosting gamma-ray binaries. Inclined disc models have also been proposed for Be/X-ray binary systems \citep[see e.g.][and references therein]{brown19}. The reason for this could be that the systems are either too young or that the compact object spends too little time near the optical star to allow tidal forces to align the orbit of a compact source. 

%%%%%%%%%%%%%%%%%%%%%%%%%%%%%%%%%%%%%%
%%%%%%%%%%%%%%%%%%%%%%%%%%%%%%%%%%%%%% 
%%%%%%%%%%%%%%%%%%%%%%%%%%%%%%%%%%%%%%
\begin{figure*}
\includegraphics[width=0.48\linewidth]{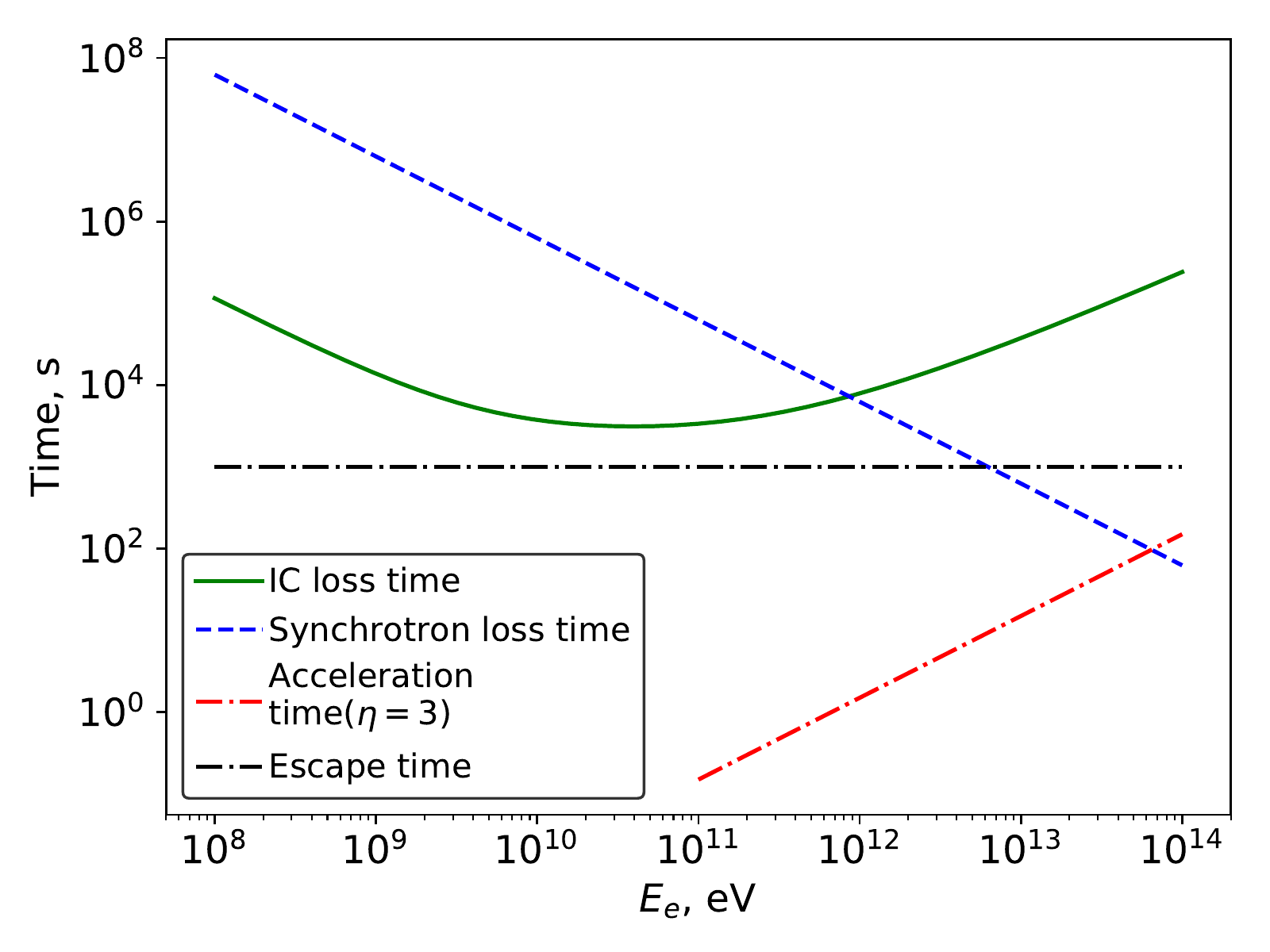}
\includegraphics[width=0.48\linewidth]{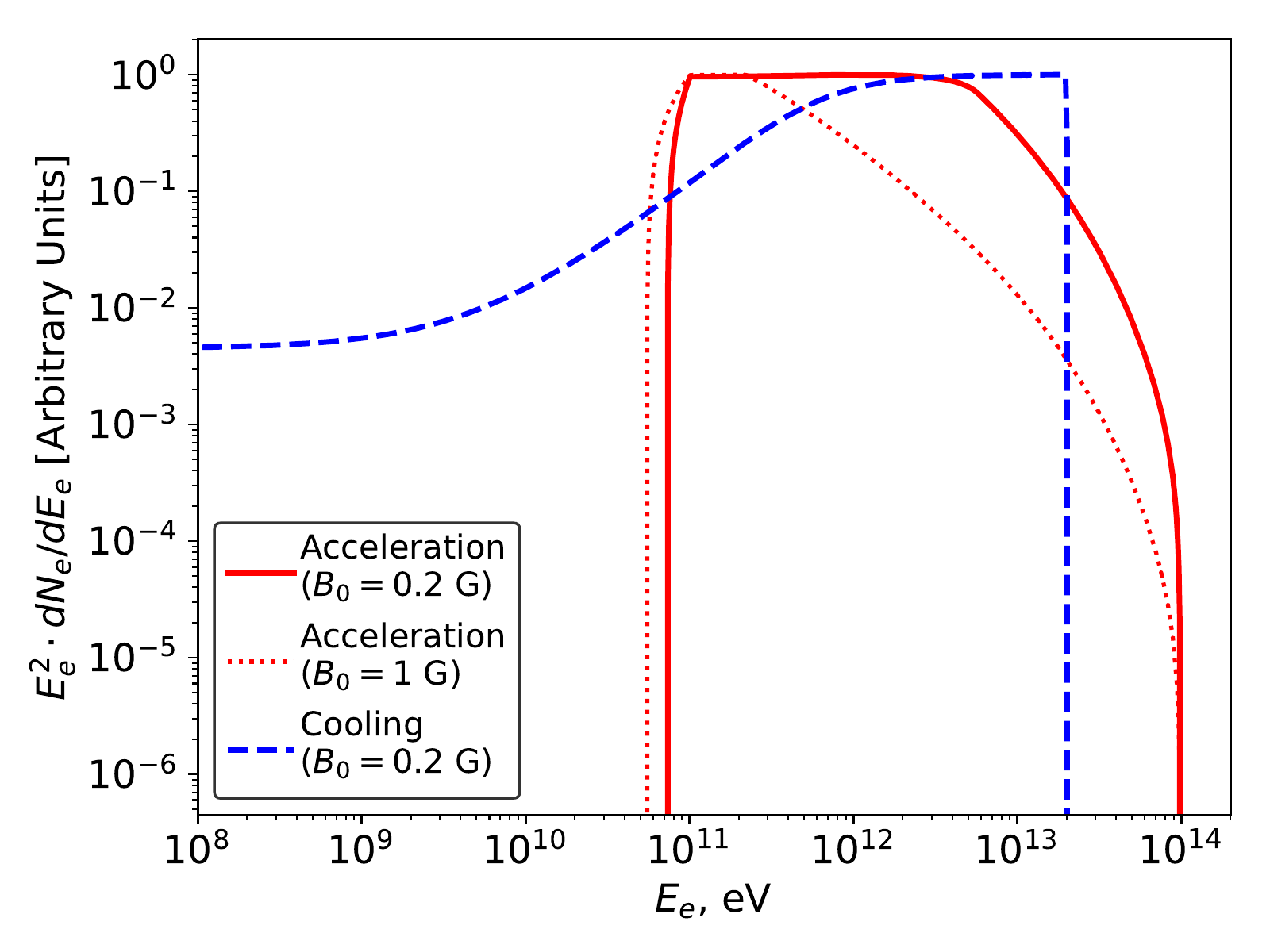}
\caption{Left: Synchrotron and IC cooling times along with characteristic acceleration timescale for electrons. Right: Spectra of electrons produced by the ``cooling'' and ``acceleration'' scenarios. }
\label{fig:tcool}
\end{figure*}
%%%%%%%%%%%%%%%%%%%%%%%%%%%%%%%%%%%%%%
\subsubsection{Spectrum formation}
\label{sec:spectrum_formation}
Similar to previous works (see e.g. \cite{takata17})  we propose that the X-ray and TeV emission observed from \psrj  originates from synchrotron and inverse Compton (IC) 
processes, respectively. The X-ray photon slope of $\Gamma\sim1.5$ allows the slope of corresponding electron spectrum to be estimated as $\Gamma_e\sim2$. At the same time the level of unpulsed radio emission reported by~\citet{ng19} does not allow a simple continuation of $\Gamma\sim 1.5$ from radio up to X-ray energies and thus suggests the presence of a low-energy feature (break or a cut-off) in the electron spectrum.
In what follows below we discuss two possible mechanisms for the formation of such a spectrum, namely the cooling and acceleration scenarios.

\noindent\textit{\textbf{Cooling scenario:}} close to mono-energetic electrons ($E_e\sim 20$~TeV) are continuously injecting by the pulsar wind at the pulsar-stellar wind interaction region. Continuous synchrotron and IC losses lead to the formation of the modified electrons spectrum and subsequently the observed X-ray and TeV emission.

\noindent\textit{\textbf{Acceleration scenario:}} similarly to the \emph{cooling scenario} mono-energetic electrons are injected, but the energy is much lower % than in the  ``cooling scenario'' 
($E_e\sim 0.1$~TeV). The injected electrons accelerate at the pulsar-stellar winds interaction shock on a characteristic timescale 
\begin{equation}
t_{\rm acc} \approx 0.1\left(E_e/1\,\mbox{TeV}\right)\eta (B_0/1\,\mbox{G})^{-1}\quad\mbox{s}
\label{eq:t_acc}    
\end{equation}
where $B_0$ is a magnetic field at the shock and $\eta\geq 1$ is the acceleration efficiency~\citep[see e.g.][for the details]{khangulian08}.

For typical magnetic field values considered for gamma-ray binaries, this timescale is much shorter than the synchrotron and IC loss timescales for the electrons with energies $\lesssim 100$~TeV. The acceleration of the electrons can lead to the fast formation of a %cut-off powerlaw spectrum of electrons 
power-law (with a cut-off) electron spectrum 
with a slope of $\Gamma_e \sim -2$ (typical for Fermi acceleration) and a cut-off at $\sim 30$~TeV energies. The accelerated particles escape from the system on a timescale of $t_{esc}\gtrsim d_*/c\sim 500$~s. 
%On longer than acceleration scales the electrons' spectrum is modified by synchrotron and IC losses which lead to the observed X-ray/TeV emission.
This approach is similar to the one used by \cite{takata17}.

% To calculate the spectrum of electrons in cooling and acceleration scenarii we developed the code which calculates the change of continuously injected spectrum of electrons by the IC and synchrotron losses. The electrons are assumed to be injected %for the escape time scale 
% with a monochromatic (``cooling scenario'') or a cut-off powerlaw (``acceleration scenario'') spectra.
% The corresponding modified by IC and synchrotron losses spectra of electrons for ``cooling'' and ``acceleration'' scenarii are shown in Fig.~\ref{fig:tcool}, right panel. 

The left panel of Fig.~\ref{fig:tcool} illustrates the cooling times for electrons due to synchrotron and IC (using the approximation of~\citealt{khangulian14}) emission for the parameters typical for Be-star gamma-ray binary systems (magnetic field $B_0=0.2$~G; distance from the star to the emission region $d_{*}=0.7$~au). The red dot-dashed line shows the acceleration time (Eq.~\ref{eq:t_acc}) and horizontal black dot-dashed line shows the  escape time from the system (selected to be of the order of $10^3$~s), while the dash blue and solid green lines show the synchrotron and IC cooling times, respectively.  For these representative parameters it can be seen that the synchrotron and IC cooling timescales are longer than the escaping time at low energy ($E_e<5\times10^{12}$eV) and do not modify the electron's spectrum. However, at higher energies the synchrotron cooling time  is short enough to substantially modify the injected spectrum (see the right panel of Fig. \ref{fig:tcool}).

The electron spectrum for both the cooling and acceleration scenarios is determined by numerically calculating the synchrotron and IC losses of a continuously injected spectrum of electrons, until a steady solution is obtained.  The electrons are assumed to be injected with a mono-energetic (cooling scenario) or a power-law with an exponential cutoff (acceleration scenario) spectra.  The resulting electron spectra are shown in the right panel of  Fig.~\ref{fig:tcool}.

% Left panel of Fig.~\ref{fig:tcool} illustrates cooling times for electrons due to synchrotron and IC (using the approximation of~\citealt{khangulian14}) emission for the parameters typical for Be-star binary systems (magnetic field $B_0=0.2$~G; distance from the star to the emission region $d_{*}=0.7$~au). Red dot-dashed line shows the acceleration time (Eq.~\ref{eq:t_acc}) and horizontal black dot-dashed line stands for the escape time from the system selected to be of order of $10^3$~s. One can see that for these parameters IC and synchrotron cooling are longer than escaping time at low energy ($E<5\times10^{12}$eV) and do not modify the electron's spectrum. At higher energies however the synchrotron cooling time  is short enough to substantially modify the injected spectrum, see right panel of Fig. \ref{fig:tcool}

\subsubsection{Spectral modelling}
%\todo[inline]{Add some references to optics section here}
As the pulsar start to approach to periastron ($t_{\rm per}-600$\,d) the interaction of the pulsar wind with the wind of the Be star shifts the emission region towards the pulsar. The subsequent increase of the magnetic field in the emission region naturally explains the observed rise of the X-ray flux.

%The X-ray flux first increases as the compact object approaches the periastron which can be understood if the wind-wind interaction region shifts towards the surface of the pulsar where higher magnetic fields can be naturally expected.  %Still, the density of the Be star disc (and the magnetic field in the interaction region) remains high enough to guarantee the effective cooling of the pulsar wind electrons.

The optical observations clearly demonstrates the influence %of the presence 
of the pulsar on the state of the Be-star disc. The interaction of the pulsar wind with the disc of the Be star leads to the %following 
rise of the X-ray emission. At $t_{\rm per}-160$\,d the equivalent width of the H\,$\alpha$ line reaches its maximum and at around the same time the X-ray photon index becomes equal to $\Gamma=1.5$.
%spectrum of X-ray emission becomes equal to 1.5. 
Such a value can be expected in either the case of acceleration in a strong shock (acceleration scenario) \citep[e.g.][]{1994ApJS...90..561J,2013APh....43...56B} or effective cooling (cooling scenario) \citep[e.g.][]{1970RvMP...42..237B}. This slope value remains almost constant up to hundred days after periastron. 

The rapid strong decrease of the flux close to periastron ($t_{\rm per}-20$ to $t_{\rm per}+10$\, d) % (t$_p$-20 -- t$_p$+10) 
corresponds to the pulsar entering a sparser regions due to the inclined Be-star disc. This consequently shifts the wind-wind interaction region further away from the pulsar and decreases the magnetic field in the emission region.
The shift of the emission region closer to the star can also explain the evolution of the observed TeV emission. This emission is produced by the IC mechanism from the same population of electrons which form the X-ray component of the observed spectrum. Contrary to X-ray light curve there is no dip at  periastron, but it rather demonstrates an increase in flux. This is in line with the proposed model, as the shift of the emission region towards the Be star implies an increase in the density of the soft photons and consequently an increased level of TeV emission reported by~\citet{psrj_tev}. The drop of the TeV flux immediately after the periastron is roughly coincident with second X-ray maximum and can be attributed to the attenuation of Be star's photons by dense regions of the disk. 

\cite{takata17} studied details of the X-ray and TeV light curves in the scenario similar to the acceleration one. This work was done before the actual periastron passage took place and lacked knowledge of system parameters. This model didn't consider the possibility of the disc inclination and attributed two peaks in the X-ray light curve to the interplay between the orbital dependence of magnetisation parameter and Doppler boosting. Subsequent observations demonstrated that the model should be adjusted to explain the observed shape and positions of X-ray peaks.

The GeV emission observed by \flat is stable along the whole orbit (see Fig.~\ref{fig:lat_lc}) and in our model this is produced by the pulsed magnetospheric emission from the pulsar. The quality of the data doesn't allow to observe the variability of the GeV emission predicted by \cite{takata17}. 

\begin{figure}
\includegraphics[width=\linewidth]{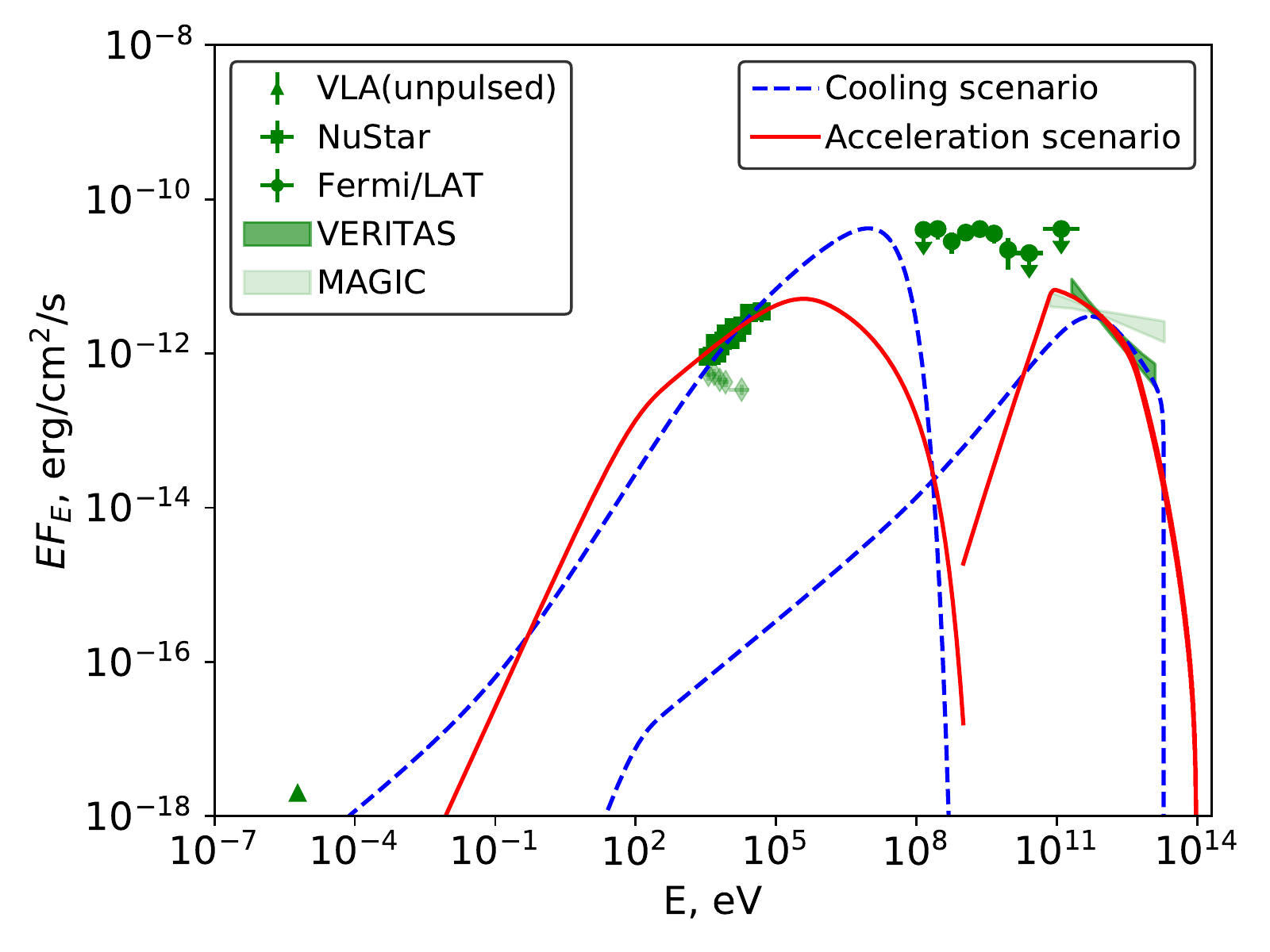}
\caption{The radio to TeV spectrum of \psrj. The green  symbols illustrate the spectrum of the source close to periastron: VLA(triangles) \flat (circles), \nus(squares), VERITAS and MAGIC(semi-transparent regions). The green diamonds  show the \nus spectrum of \psrj measured at $\sim -400$~days before to periastron. The curves show the  results of the broad-band spectral modelling of the system for the ``cooling'' (blue dashed) and acceleration (red solid) scenarios; see text for details. The radio and TeV data are from~\citet{ng19, psrj_tev}. }
\label{fig:spectrum}
\end{figure}

The X-ray to TeV SED of the system is shown in Fig.~\ref{fig:spectrum} along with the results of spectral modelling of the system. Green symbols illustrate the available data close to periastron: VLA (triangles; \citet{ng19}), \nus(squares), \flat(circles), VERITAS and MAGIC(semi-transparent areas; \citet{psrj_tev}). The green semi-transparent diamond points show the\nus spectrum taken at $\sim -400$~days before the periastron passage. The solid and dashed curves present the results of broad-band spectral modelling of the system for the ``cooling'' and ``acceleration'' scenarios. 

% The spectra of photons produced by considered population of electrons were calculated with \texttt{naima v.0.8.3} package \citep{naima}, which uses cross-sections and SED analytic approximations for the IC and synchrotron emission by~\citet{aharonian81,aharonian10,khangulian14}.

The synchrotron and IC emission was calculated with the \texttt{naima v.0.8.3} package \citep{naima}, which uses the analytic approximations for the IC and synchrotron emission developed by~\citet{aharonian81,aharonian10,khangulian14}.
The solid red line in Fig.~\ref{fig:spectrum} shows the result for the ``acceleration'' scenario and corresponds to an emitting region located at a mean distance of  $d=0.7$~au from the star, a magnetic field strength of $B_0=0.2$~G and an escape time from the system of $t_{esc}=10^3$~s.
The spectral modelling of the  ``cooling'' scenario (blue dashed curves) for the same location of the emitting region requires a slightly higher magnetic field ($B_0=1$~G), and does not suggest the escape of electrons from the system. For both scenarios we assume the star to have a luminosity of  $L_*=6\cdot 10^{37}$~erg/s and a temperature $T_*=3\cdot 10^4$~K. The presented spectra directly correspond to the electron spectra shown in right panel of Fig.~\ref{fig:tcool}.

%%%%%%%%%%%%%%%%%%%%%%%%%%%%%%%%%%%%%%
\subsection{Comparison to other ``inclined disc'' systems}

% Despite ``inclined disc'' geometry of \psrj is suggested to be similar to several other gamma-ray binaries (e.g. \psrb or \hessj), the behaviour of this system during the periastron passage does not exactly resembles the behaviour of these systems. 

While the  ``inclined disc'' geometry proposed for \psrj would suggest that it should be very similar to other gamma-ray binaries (e.g. \psrb or \hessj), the system's behaviour during the periastron passage does not exactly resembles the behaviour of these other systems. 
While, similar to \psrb and \hessj, \psrj demonstrates a double-peaked X-ray lightcurve with a deep minimum at periastron, contrary to these systems there is no hints of orbital variations of the hydrogen column density. 
%are seen in \psrj. 
The orbital X-ray spectral variability (which shows significant spectral hardening close to periastron) is also different from \psrb \citep[softening of the spectrum close to periastron,][] {psrb2014} and \hessj \citep[constant slope,][]{we_hessj}.

In the optical band, the differences in the behaviour of the H\,$\alpha$ emission line,  between \psrj and \psrb systems are clearly seen.  In contrast to what is observed in PSR B1259-63 \citep[see][]{psrb2014} where the EW (which is linked to the disc size) increases as the neutron star approaches periastron, our data set, and that of \cite{coe19},  %show that the disc reached its maximum size %about 100 days before periastron passage. 
shows the EW reaches a maximum about 180\,d  before periastron. 
But afterwards, in response to the proximity of the compact companion the disc size decreases again to levels similar to those shown in \citet{coe19} at epochs earlier than MJD$\sim$57700 (a few hundreds of days before periastron). After the periastron passage, all the H\,$\alpha$ parameters show fluctuations, and after $\sim$50 days they slowly and smoothly recover to reach values similar to those before periastron. We interpret this as disruptions caused by the tidal interaction with the compact object. Another difference is that while,  for \psrb,  the increase in the equivalent width is steeper than the decrease, in \psrj we find the opposite behaviour.  Both \psrb and \psrj  show evidence of a tilted disc with respect to the orbital plane. The main difference between them is that the former one has a bigger, denser,  and very stable disc  ~\citep{psrb2014} while disc in the latter is smaller and unstable \citep{coe19}. Since in both cases the pulsar crosses the disc near periastron the most dramatic effects seen in \psrj can, therefore, be interpreted as a consequence of the properties of its circumstellar disc.
  A comparison between how the pulsar disrupts a lower density and a higher density circumstellar disc can be seen in the SPH  modelling shown in fig.~1 \& 2 of \citet{takata12}. There the authors compare, for \psrb, the pulsar approaching the star/circumstellar disc for a base density of $\rho_0 = 10^{-11}$ and $10^{-9}$\,g\,cm$^{-3}$.

%In \psrj $H\alpha$ line seems to be significantly narrower ($-EW\lesssim -15$ vs. $-EW\lesssim -80$ for \psrb). The orbital evolution of the $-EW$ for \psrj is characterized by a double-peak structure with a local minimum close to periastron, while for \psrb system $-EW$ demonstrates only one peak along the orbit, with a maximum at periastron.
%\todo[inline]{Explain the reason of a single peak}
%\todo[inline]{In 1259 it's multiple lines blended into each other. You can see the shape of the line shifting and the line is clearly a blend of different lines. If you look at Fig 2. in https://arxiv.org/abs/1511.01224 the spectra we got in 2014 shows this. }

%The described differences can be attributed to different physical properties of Be-star discs and the power of the pulsar wind. In \psrb the disc seems to be dense and broad; even at close to periastron phases the pulsar remains in a relatively dense regions. The graduate increase of $-EW$ as the pulsar approaches the periastron can be explained as an increase of the size/temperature \red{please check!} of the disc due \todo{tidal interactions!} to continuous pumping of pulsar wind energy. This process continues until post-periastron destroy of the disc (rapid decrease of $-EW$). The disc of Be star in \psrj system is smaller and sparser.

The maxima of the (absolute value) equivalent width of the H\,$\alpha$ line are offset from the peaks of X-ray lightcurve (see Fig.~\ref{fig:xrt_lc_index}, right panel). Assuming that the X-ray flux maxima correspond to the moments of the plane of the disc-crossing, we conclude that the disc has a maximal size \citep[(see e.g.][]{grundstrom06}) significantly before and after disc crossing events. This can be understood, if a few competing processes take place in this system, e.g. star-pulsar tidal interactions (increasing disc size and build-up rate, as the pulsar approaches the periastron) and perturbations of the disc by the pulsar wind (continuous destroyment of the disc which leads to gradual decreasing of its size, e.g. sharp cutting the disc at distances higher than pulsar-star separation, see e.g.~\citep{monageng17,coe19}). Note that in \citet{takata17} it was demonstrated that if the base density is less than
$\rho_0 < 10^{-10}$ g/cm$^3$, the pulsar wind can strip off an outer part of the Be disk, truncating it at a radius smaller than the pulsar orbit, 
and only the post-periastron X-ray flare may be related to disc crossing \citep{li18}.

The magnetic field in the emitting region typically considered for \psrb is about $\sim 1$~G~\citep[see e.g.][]{psrb2014} which is comparable to values found here within the cooling scenario and somewhat higher than what is required for the acceleration scenario ($\sim 0.2$~G). In terms of the energetics the spin-down luminosity of \psrb exceeds that of \psrj by a factor of 5 \citep{romoli17, psrj_tev}, while the spatial scales of both systems at periastron are comparable ($\sim 1$~au). 
The higher magnetic field in \psrb allows fast and more efficient cooling of the pulsar wind's electrons via synchrotron losses at timescales shorter or comparable to \psrj. Within the cooling scenario this should lead to a similar formation of a low-energy tail in the spectrum of electrons. The characteristic tail's slope of $\Gamma_e = -2$ should result in an X-rays slope of $\Gamma \sim -1.5$ during the periods of disc crossing. Contrary, significant spectral softening (up to photon X-ray slope $\sim -2$) was observed during these periods~\citep{psrb2014}. Such a discrepancy is challenging  to explain within the ``cooling scenario'', however it could be understood within ``acceleration scenaio''.

% Fig.~\ref{fig:tcool}, right panel, illustrates changes in modified by synchrotron/IC cooling spectrum of initially cut-off powerlaw (with the slope of -2) spectrum of accelerated electrons with respect to different values of the magnetic field in the emitting region. 
Fig.~\ref{fig:tcool} (right panel) illustrates how different values of the magnetic field in the emitting region modify the final electron spectrum, for an initially injected cut-off powerlaw (with the slope of $-2$) electron spectrum in the "acceleration scenario". Namely, the increased value of the magnetic field leads to a significant softening of the spectrum at energies comparable to the injection energy ($\sim 0.1$~TeV). The electron spectrum is modified in a similar way for a constant magnetic field but for longer electron escape times. % from the system.

Within the ``acceleration scenario'' significant softening of the electrons' spectrum (and consequent softening of X-rays) during the disc crossing in \psrb can be understood if the emitting region approaches the pulsar (magnetic field in the region increases) and/or escape time from the system significantly increase when the pulsar enters dense disc regions. 
The weaker H\,$\alpha$ emission line observed in \psrj suggests that the disc is smaller than in in \psrb.  This means that 
% Contrary to \psrb, \psrj 
during the whole periastron passage, \psrj remains in a lower-density regions (than in \psrb) which results in shorter escape times % (indicated by lower $-EW$ $H\alpha$ values; resulting in shorter escape times) 
and the magnetic field in the emitting region remains relatively low. This results in a hard X-ray slope of \psrj system during the whole periastron passage.

\section{Conclusions}
In this paper we analyse the results of multi-wavelength observation of \psrj as it is approaching and passing by its periastron in 2017. The observations of the H\,$\alpha$
line  show a clear variability of the Be star disc, both in the change in equivalent width, the line separation and the V/R profile. X-ray emission from the system is also highly variable, increasing as the pulsar approaches periastron, but with a sharp minimum around the periastron itself. % Clear link between the X-ray emission and the state of the Be star disc is demonstrated by the fact that, as it is follows from the change of the X-ray spectral index, the electrons of the pulsar wind enter the regime of the effective cooling / acceleration on the strong shock at the moment of the peak value of the H$_{\alpha}$ equivalent width. 
A clear link between the X-ray emission and the state of the Be star disc is demonstrated by the fact  the X-ray spectral index changes at the same  moment as the H\,$\alpha$ equivalent width peaks. 

The assumption that the disc of the Be star is inclined to the orbital plane allows the observed high energy variability  to be explained in a natural way.  In the proposed model  X-ray and TeV emission observed from \psrj  originate  from  synchrotron  and  inverse Compton  processes,  respectively. The observed spectrum from the source can be understood within the ``cooling'' or the ``acceleration'' scenario which relies on cooling/acceleration of monoenergetic electrons of pulsar wind, see Sec.~\ref{sec:spectrum_formation} for the details. While both scenarios are able to satisfactory describe the SED of \psrj system, we argue that ``acceleration scenario'' seems to be preferable for other gamma-ray binaries, e.g. \psrb and thus is natural for \psrj as well.

%We argue thus that the ``acceleration scenario'' seems to be preferable for \psrj and can play a role in other gamma-ray binaries, e.g. \psrb.

%At periastron the pulsar is out of the disc, and the emission region is located further from the pulsar than it was when the pulsar was inside the dense part of the disc. Such a shift decrease the magnetic field, and hence the X-ray emission has a local minimum at periastron. At the same time shift of the emission region closer to the Be star increase the density of the soft photons  at periastron, and correspondingly the flux of the TeV emission. We discuss two possible mechanisms of the 

%Observations demonstrate rise of both X-ray emission and H$_{\alpha}$ equivalent width starting at least 600 days before the periastron. \todo{in X-ray actually the rise started much earlier, and in optics I am not sure that there are data before ones published in Coe. We should think better how to formulate this, or should we start from later stage already?} 

%\newline
% \noindent\textit{Acknowledgements}. 
\section*{Acknowledgements}
The authors acknowledge support by the state of Baden-W\"urttemberg through bwHPC. 
This work was supported by DFG through the grant MA 7807/2-1, partially supported by the Spanish Government under grant PGC2018-093741-B-C21 (MICIU/AEI/FEDER, UE), and the Russian Foundation for Basic Research project 17-52-80139 BRICS-a (SST).
This work made use of data supplied by the UK Swift Science Data Centre at the University of Leicester. We acknowledge the use of public data from the Swift data archive and thanks the entire Swift team for accepting and planning Target-of-Opportunity requests.
We also acknowledge funding to support our NOT observations from the Finnish Centre for Astronomy with ESO (FINCA), University of Turku, Finland (Academy of Finland grant nr 306531). The authors wish to acknowledge the DJEI/DES/SFI/HEA Irish Centre for High-End Computing (ICHEC) for the provision of computational facilities and support. We would also like to acknowledge networking support by the COST Actions CA16214 and CA16104.

%\todo[inline]{Is there a chance to distinguish between cooling and acceleration? }

%\todo[inline]{How can we check our model? Clear predictions for the next periastron passage: in optics/X-rays/GeV/TeV}

%\begin{figure}
%    \centering
%    \includegraphics[width=0.99\linewidth%]{x_ray_optics.pdf}
%    \caption{X-ray and optics data}
%    \label{fig:optics_xray}
%\end{figure}

\bibliographystyle{mn2e}
% Bibliography and bibfile
\def\aj{AJ}%
          % Astronomical Journal
\def\actaa{Acta Astron.}%
          % Acta Astronomica
\def\araa{ARA\&A}%
          % Annual Review of Astron and Astrophys
\def\apj{ApJ}%
          % Astrophysical Journal
\def\apjl{ApJ}%
          % Astrophysical Journal, Letters
\def\apjs{ApJS}%
          % Astrophysical Journal, Supplement
\def\ao{Appl.~Opt.}%
          % Applied Optics
\def\apss{Ap\&SS}%
          % Astrophysics and Space Science
\def\aap{A\&A}%
          % Astronomy and Astrophysics
\def\aapr{A\&A~Rev.}%
          % Astronomy and Astrophysics Reviews
\def\aaps{A\&AS}%
          % Astronomy and Astrophysics, Supplement
\def\azh{AZh}%
          % Astronomicheskii Zhurnal
\def\baas{BAAS}%
          % Bulletin of the AAS
\def\bac{Bull. astr. Inst. Czechosl.}%
          % Bulletin of the Astronomical Institutes of Czechoslovakia
\def\caa{Chinese Astron. Astrophys.}%
          % Chinese Astronomy and Astrophysics
\def\cjaa{Chinese J. Astron. Astrophys.}%
          % Chinese Journal of Astronomy and Astrophysics
\def\icarus{Icarus}%
          % Icarus
\def\jcap{J. Cosmology Astropart. Phys.}%
          % Journal of Cosmology and Astroparticle Physics
\def\jrasc{JRASC}%
          % Journal of the RAS of Canada
\def\mnras{MNRAS}%
          % Monthly Notices of the RAS
\def\memras{MmRAS}%
          % Memoirs of the RAS
\def\na{New A}%
          % New Astronomy
\def\nar{New A Rev.}%
          % New Astronomy Review
\def\pasa{PASA}%
          % Publications of the Astron. Soc. of Australia
\def\pra{Phys.~Rev.~A}%
          % Physical Review A: General Physics
\def\prb{Phys.~Rev.~B}%
          % Physical Review B: Solid State
\def\prc{Phys.~Rev.~C}%
          % Physical Review C
\def\prd{Phys.~Rev.~D}%
          % Physical Review D
\def\pre{Phys.~Rev.~E}%
          % Physical Review E
\def\prl{Phys.~Rev.~Lett.}%
          % Physical Review Letters
\def\pasp{PASP}%
          % Publications of the ASP
\def\pasj{PASJ}%
          % Publications of the ASJ
\def\qjras{QJRAS}%
          % Quarterly Journal of the RAS
\def\rmxaa{Rev. Mexicana Astron. Astrofis.}%
          % Revista Mexicana de Astronomia y Astrofisica
\def\skytel{S\&T}%
          % Sky and Telescope
\def\solphys{Sol.~Phys.}%
          % Solar Physics
\def\sovast{Soviet~Ast.}%
          % Soviet Astronomy
\def\ssr{Space~Sci.~Rev.}%
          % Space Science Reviews
\def\zap{ZAp}%
          % Zeitschrift fuer Astrophysik
\def\nat{Nature}%
          % Nature
\def\iaucirc{IAU~Circ.}%
          % IAU Cirulars
\def\aplett{Astrophys.~Lett.}%
          % Astrophysics Letters
\def\apspr{Astrophys.~Space~Phys.~Res.}%
          % Astrophysics Space Physics Research
\def\bain{Bull.~Astron.~Inst.~Netherlands}%
          % Bulletin Astronomical Institute of the Netherlands
\def\fcp{Fund.~Cosmic~Phys.}%
          % Fundamental Cosmic Physics
\def\gca{Geochim.~Cosmochim.~Acta}%
          % Geochimica Cosmochimica Acta
\def\grl{Geophys.~Res.~Lett.}%
          % Geophysics Research Letters
\def\jcp{J.~Chem.~Phys.}%
          % Journal of Chemical Physics
\def\jgr{J.~Geophys.~Res.}%
          % Journal of Geophysics Research
\def\jqsrt{J.~Quant.~Spec.~Radiat.~Transf.}%
          % Journal of Quantitiative Spectroscopy and Radiative Trasfer
\def\memsai{Mem.~Soc.~Astron.~Italiana}%
          % Mem. Societa Astronomica Italiana
\def\nphysa{Nucl.~Phys.~A}%
          % Nuclear Physics A
\def\physrep{Phys.~Rep.}%
          % Physics Reports
\def\physscr{Phys.~Scr}%
          % Physica Scripta
\def\planss{Planet.~Space~Sci.}%
          % Planetary Space Science
\def\procspie{Proc.~SPIE}%
          % Proceedings of the SPIE
\let\astap=\aap
\let\apjlett=\apjl
\let\apjsupp=\apjs
\let\applopt=\ao
%\bibliography{bibliography}

\begin{thebibliography}{47}
\expandafter\ifx\csname natexlab\endcsname\relax\def\natexlab#1{#1}\fi

\bibitem[{{Abdo} {et~al}\mbox{.}(2009){Abdo}, {Ackermann}, {Ajello},
  {Anderson}, {Atwood}, {Axelsson}, {Baldini}, {Ballet}, {Barbiellini},
  {Baring}, {Bastieri}, {Baughman}, {Bechtol}, {Bellazzini}, {Berenji},
  {Bignami}, {Blandford}, {Bloom}, {Bonamente}, {Borgland}, {Bregeon}, {Brez},
  {Brigida}, {Bruel}, {Burnett}, {Caliandro}, {Cameron}, {Caraveo},
  {Casandjian}, {Cecchi}, {{\c{C}}elik}, {Chekhtman}, {Cheung}, {Chiang},
  {Ciprini}, {Claus}, {Cohen-Tanugi}, {Conrad}, {Cutini}, {Dermer}, {de
  Angelis}, {de Luca}, {de Palma}, {Digel}, {Dormody}, {do Couto e Silva},
  {Drell}, {Dubois}, {Dumora}, {Farnier}, {Favuzzi}, {Fegan}, {Fukazawa},
  {Funk}, {Fusco}, {Gargano}, {Gasparrini}, {Gehrels}, {Germani}, {Giebels},
  {Giglietto}, {Giommi}, {Giordano}, {Glanzman}, {Godfrey}, {Grenier},
  {Grondin}, {Grove}, {Guillemot}, {Guiriec}, {Gwon}, {Hanabata}, {Harding},
  {Hayashida}, {Hays}, {Hughes}, {J{\'o}hannesson}, {Johnson}, {Johnson},
  {Johnson}, {Kamae}, {Katagiri}, {Kataoka}, {Kawai}, {Kerr}, {Kn{\"o}dlseder},
  {Kocian}, {Kuss}, {Land e}, {Latronico}, {Lemoine-Goumard}, {Longo},
  {Loparco}, {Lott}, {Lovellette}, {Lubrano}, {Madejski}, {Makeev}, {Marelli},
  {Mazziotta}, {McConville}, {McEnery}, {Meurer}, {Michelson}, {Mitthumsiri},
  {Mizuno}, {Monte}, {Monzani}, {Morselli}, {Moskalenko}, {Murgia}, {Nolan},
  {Norris}, {Nuss}, {Ohsugi}, {Omodei}, {Orlando}, {Ormes}, {Paneque},
  {Parent}, {Pelassa}, {Pepe}, {Pesce-Rollins}, {Pierbattista}, {Piron},
  {Porter}, {Primack}, {Rain{\`o}}, {Rando}, {Ray}, {Razzano}, {Rea}, {Reimer},
  {Reimer}, {Reposeur}, {Ritz}, {Rochester}, {Rodriguez}, {Romani}, {Ryde},
  {Sadrozinski}, {Sanchez}, {Sander}, {Parkinson}, {Scargle}, {Sgr{\`o}},
  {Siskind}, {Smith}, {Smith}, {Spand re}, {Spinelli}, {Starck}, {Strickman},
  {Suson}, {Tajima}, {Takahashi}, {Takahashi}, {Tanaka}, {Thayer}, {Thompson},
  {Tibaldo}, {Tibolla}, {Torres}, {Tosti}, {Tramacere}, {Uchiyama}, {Usher},
  {Van Etten}, {Vasileiou}, {Vilchez}, {Vitale}, {Waite}, {Wang}, {Watters},
  {Winer}, {Wolff}, {Wood}, {Ylinen}, {Ziegler}, \& {Fermi LAT
  Collaboration}}]{fermipulsars09}
{Abdo} A.~A. {et~al.}, 2009, Science, 325, 840

\bibitem[{{Abeysekara} {et~al}\mbox{.}(2018){Abeysekara}, {Benbow}, {Bird},
  {Brill}, {Brose}, {Buckley}, {Chromey}, {Daniel}, {Falcone}, {Finley}, \&
  et~al.}]{psrj_tev}
{Abeysekara} A.~U. {et~al.}, 2018, \apjl, 867, L19

\bibitem[{{Aharonian} {et~al}\mbox{.}(2005){Aharonian}, {Akhperjanian},
  {Beilicke}, {Bernl{\"o}hr}, {B{\"o}rst}, {Bojahr}, {Bolz}, {Coarasa},
  {Contreras}, {Cortina}, {Denninghoff}, {Fonseca}, {Girma}, {G{\"o}tting},
  {Heinzelmann}, {Hermann}, {Heusler}, {Hofmann}, {Horns}, {Jung}, {Kankanyan},
  {Kestel}, {Kohnle}, {Konopelko}, {Kranich}, {Lampeitl}, {Lopez}, {Lorenz},
  {Lucarelli}, {Mang}, {Mazin}, {Meyer}, {Mirzoyan}, {Moralejo},
  {O{\~n}a-Wilhelmi}, {Panter}, {Plyasheshnikov}, {P{\"u}hlhofer}, {de los
  Reyes}, {Rhode}, {Ripken}, {Rowell}, {Sahakian}, {Samorski}, {Schilling},
  {Siems}, {Sobzynska}, {Stamm}, {Tluczykont}, {Vitale}, {V{\"o}lk}, {Wiedner},
  \& {Wittek}}]{hegra05}
{Aharonian} F. {et~al.}, 2005, \aap, 431, 197

\bibitem[{{Aharonian} \& {Atoyan}(1981)}]{aharonian81}
{Aharonian} F.~A., {Atoyan} A.~M., 1981, \apss, 79, 321

\bibitem[{{Aharonian}, {Kelner} \& {Prosekin}(2010){Aharonian}, {Kelner}, \&
  {Prosekin}}]{aharonian10}
{Aharonian} F.~A., {Kelner} S.~R., {Prosekin} A.~Y., 2010, \prd, 82, 043002

\bibitem[{{Aliu} {et~al}\mbox{.}(2014){Aliu}, {Aune}, {Behera}, {Beilicke},
  {Benbow}, {Berger}, {Bird}, {Buckley}, {Bugaev}, {Cardenzana}, {Cerruti},
  {Chen}, {Ciupik}, {Connolly}, {Cui}, {Duke}, {Dumm}, {Errando}, {Falcone},
  {Federici}, {Feng}, {Finley}, {Fortin}, {Fortson}, {Furniss}, {Galante},
  {Gillanders}, {Griffin}, {Griffiths}, {Grube}, {Gyuk}, {Hanna}, {Holder},
  {Hughes}, {Humensky}, {Kaaret}, {Kargaltsev}, {Kertzman}, {Khassen}, {Kieda},
  {Krawczynski}, {Lang}, {Madhavan}, {Maier}, {Majumdar}, {McCann}, {Moriarty},
  {Mukherjee}, {Nieto}, {O'Faol{\'a}in de Bhr{\'o}ithe}, {Ong}, {Otte},
  {Pandel}, {Perkins}, {Pohl}, {Popkow}, {Prokoph}, {Quinn}, {Ragan},
  {Rajotte}, {Reyes}, {Reynolds}, {Richards}, {Roache}, {Sembroski}, {Skole},
  {Staszak}, {Telezhinsky}, {Theiling}, {Tucci}, {Tyler}, {Varlotta},
  {Vincent}, {Wakely}, {Weekes}, {Weinstein}, {Welsing}, {Williams}, \&
  {Zitzer}}]{aliu14}
{Aliu} E. {et~al.}, 2014, \apj, 783, 16

\bibitem[{{Bednarek}, {Banasi{\'n}ski} \& {Sitarek}(2018){Bednarek},
  {Banasi{\'n}ski}, \& {Sitarek}}]{bednarek18}
{Bednarek} W., {Banasi{\'n}ski} P., {Sitarek} J., 2018, Journal of Physics G
  Nuclear Physics, 45, 015201

\bibitem[{{Bell}(2013)}]{2013APh....43...56B}
{Bell} A.~R., 2013, Astroparticle Physics, 43, 56

\bibitem[{{Blumenthal} \& {Gould}(1970)}]{1970RvMP...42..237B}
{Blumenthal} G.~R., {Gould} R.~J., 1970, Reviews of Modern Physics, 42, 237

\bibitem[{{Brown} {et~al}\mbox{.}(2019){Brown}, {Coe}, {Ho}, \&
  {Okazaki}}]{brown19}
{Brown} R.~O., {Coe} M.~J., {Ho} W.~C.~G., {Okazaki} A.~T., 2019, \mnras, 488,
  387

\bibitem[{{Camilo} {et~al}\mbox{.}(2009){Camilo}, {Ray}, {Ransom}, {Burgay},
  {Johnson}, {Kerr}, {Gotthelf}, {Halpern}, {Reynolds}, {Romani}, {Demorest},
  {Johnston}, {van Straten}, {Saz Parkinson}, {Ziegler}, {Dormody}, {Thompson},
  {Smith}, {Harding}, {Abdo}, {Crawford}, {Freire}, {Keith}, {Kramer},
  {Roberts}, {Weltevrede}, \& {Wood}}]{Camilo09}
{Camilo} F. {et~al.}, 2009, \apj, 705, 1

\bibitem[{{Chernyakova} {et~al}\mbox{.}(2019){Chernyakova}, {Malyshev},
  {Paizis}, {La Palombara}, {Balbo}, {Walter}, {Hnatyk}, {van Soelen},
  {Romano}, {Munar-Adrover}, {Vovk}, {Piano}, {Capitanio},
  {Falceta-Gon{\c{c}}alves}, {Landoni}, {Luque-Escamilla}, {Mart{\'\i}},
  {Paredes}, {Rib{\'o}}, {Safi-Harb}, {Saha}, {Sidoli}, \&
  {Vercellone}}]{chernyakova19}
{Chernyakova} M. {et~al.}, 2019, \aap, 631, A177

\bibitem[{{Chernyakova} {et~al}\mbox{.}(2006){Chernyakova}, {Neronov},
  {Lutovinov}, {Rodriguez}, \& {Johnston}}]{chernyakova06}
{Chernyakova} M., {Neronov} A., {Lutovinov} A., {Rodriguez} J., {Johnston} S.,
  2006, \mnras, 367, 1201

\bibitem[{{Chernyakova} {et~al}\mbox{.}(2015){Chernyakova}, {Neronov}, {van
  Soelen}, {Callanan}, {O'Shaughnessy}, {Babyk}, {Tsygankov}, {Vovk},
  {Krivonos}, {Tomsick}, {Malyshev}, {Li}, {Wood}, {Torres}, {Zhang},
  {Kretschmar}, {McSwain}, {Buckley}, \& {Koen}}]{psrb2014}
{Chernyakova} M. {et~al.}, 2015, \mnras, 454, 1358

\bibitem[{{Coe} {et~al}\mbox{.}(2019){Coe}, {Okazaki}, {Steele}, {Ng}, {Ho},
  {Lyne}, {Stappers}, {Johnson}, {Ray}, \& {Kerr}}]{coe19}
{Coe} M.~J. {et~al.}, 2019, \mnras, 485, 1864

\bibitem[{{Corbet} {et~al}\mbox{.}(2019){Corbet}, {Chomiuk}, {Coe}, {Coley},
  {Dubus}, {Edwards}, {Martin}, {McBride}, {Stevens}, {Strader}, \&
  {Townsend}}]{corbet19}
{Corbet} R.~H.~D. {et~al.}, 2019, \apj, 884, 93

\bibitem[{{Dubus}(2013)}]{dubus13}
{Dubus} G., 2013, \aapr, 21, 64

\bibitem[{{Grundstrom} \& {Gies}(2006)}]{grundstrom06}
{Grundstrom} E.~D., {Gies} D.~R., 2006, \apjl, 651, L53

\bibitem[{{Hanuschik}(1989)}]{hanuschik89}
{Hanuschik} R.~W., 1989, \apss, 161, 61

\bibitem[{{Ho} {et~al}\mbox{.}(2017){Ho}, {Ng}, {Lyne}, {Stappers}, {Coe},
  {Halpern}, {Johnson}, \& {Steele}}]{ho17}
{Ho} W. C.~G., {Ng} C.~Y., {Lyne} A.~G., {Stappers} B.~W., {Coe} M.~J.,
  {Halpern} J.~P., {Johnson} T.~J., {Steele} I.~A., 2017, \mnras, 464, 1211

\bibitem[{{Hummel} \& {Hanuschik}(1997)}]{hummel97}
{Hummel} W., {Hanuschik} R.~W., 1997, \aap, 320, 852

\bibitem[{{Johnston} {et~al}\mbox{.}(1992){Johnston}, {Lyne}, {Manchester},
  {Kniffen}, {D'Amico}, {Lim}, \& {Ashworth}}]{johnston92}
{Johnston} S., {Lyne} A.~G., {Manchester} R.~N., {Kniffen} D.~A., {D'Amico} N.,
  {Lim} J., {Ashworth} M., 1992, \mnras, 255, 401

\bibitem[{{Jones}(1994)}]{1994ApJS...90..561J}
{Jones} F.~C., 1994, \apjs, 90, 561

\bibitem[{{Khangulyan}, {Aharonian} \& {Bosch-Ramon}(2008){Khangulyan},
  {Aharonian}, \& {Bosch-Ramon}}]{khangulian08}
{Khangulyan} D., {Aharonian} F., {Bosch-Ramon} V., 2008, \mnras, 383, 467

\bibitem[{{Khangulyan}, {Aharonian} \& {Kelner}(2014){Khangulyan}, {Aharonian},
  \& {Kelner}}]{khangulian14}
{Khangulyan} D., {Aharonian} F.~A., {Kelner} S.~R., 2014, \apj, 783, 100

\bibitem[{{Klement} {et~al}\mbox{.}(2017){Klement}, {Carciofi}, {Rivinius},
  {Matthews}, {Vieira}, {Ignace}, {Bjorkman}, {Mota}, {Faes}, {Bratcher},
  {Cur{\'e}}, \& {{\v{S}}tefl}}]{klement17}
{Klement} R. {et~al.}, 2017, \aap, 601, A74

\bibitem[{{Kolka} {et~al}\mbox{.}(2017){Kolka}, {Eenm{\"a}e}, {Laur}, \&
  {Aret}}]{kolka17}
{Kolka} I., {Eenm{\"a}e} T., {Laur} J., {Aret} A., 2017, Research Notes of the
  American Astronomical Society, 1, 37

\bibitem[{{Li} {et~al}\mbox{.}(2018){Li}, {Takata}, {Ng}, {Kong}, {Tam}, {Hui},
  \& {Cheng}}]{li18}
{Li} K.~L., {Takata} J., {Ng} C.~W., {Kong} A.~K.~H., {Tam} P.~H.~T., {Hui}
  C.~Y., {Cheng} K.~S., 2018, \apj, 857, 123

\bibitem[{{Lyne} {et~al}\mbox{.}(2015){Lyne}, {Stappers}, {Keith}, {Ray},
  {Kerr}, {Camilo}, \& {Johnson}}]{lyne15}
{Lyne} A.~G., {Stappers} B.~W., {Keith} M.~J., {Ray} P.~S., {Kerr} M., {Camilo}
  F., {Johnson} T.~J., 2015, \mnras, 451, 581

\bibitem[{{Malyshev} {et~al}\mbox{.}(2017){Malyshev}, {Chernyakova},
  {Santangelo}, \& {P{\"u}hlhofer}}]{we_hessj}
{Malyshev} D., {Chernyakova} M., {Santangelo} A., {P{\"u}hlhofer} G., 2017,
  arXiv e-prints

\bibitem[{{Melatos}, {Johnston} \& {Melrose}(1995){Melatos}, {Johnston}, \&
  {Melrose}}]{melatos95}
{Melatos} A., {Johnston} S., {Melrose} D.~B., 1995, \mnras, 275, 381

\bibitem[{{Monageng} {et~al}\mbox{.}(2017){Monageng}, {McBride}, {Coe},
  {Steele}, \& {Reig}}]{monageng17}
{Monageng} I.~M., {McBride} V.~A., {Coe} M.~J., {Steele} I.~A., {Reig} P.,
  2017, \mnras, 464, 572

\bibitem[{{Ng} {et~al}\mbox{.}(2019){Ng}, {Ho}, {Gotthelf}, {Halpern}, {Coe},
  {Stappers}, {Lyne}, {Wood}, \& {Kerr}}]{ng19}
{Ng} C.~Y. {et~al.}, 2019, \apj, 880, 147

\bibitem[{{Okazaki} {et~al}\mbox{.}(2002){Okazaki}, {Bate}, {Ogilvie}, \&
  {Pringle}}]{okazaki02}
{Okazaki} A.~T., {Bate} M.~R., {Ogilvie} G.~I., {Pringle} J.~E., 2002, \mnras,
  337, 967

\bibitem[{{Pal} {et~al}\mbox{.}(2019){Pal}, {Tam}, {Cui}, {Li}, {Kong}, \&
  {G{\"u}ng{\"o}r}}]{Pal19}
{Pal} P.~S., {Tam} P.~H.~T., {Cui} Y., {Li} K.~L., {Kong} A.~K.~H.,
  {G{\"u}ng{\"o}r} C., 2019, \apj, 882, 25

\bibitem[{{Panoglou} {et~al}\mbox{.}(2016){Panoglou}, {Carciofi}, {Vieira},
  {Cyr}, {Jones}, {Okazaki}, \& {Rivinius}}]{panoglou16}
{Panoglou} D., {Carciofi} A.~C., {Vieira} R.~G., {Cyr} I.~H., {Jones} C.~E.,
  {Okazaki} A.~T., {Rivinius} T., 2016, \mnras, 461, 2616

\bibitem[{{Reig} {et~al}\mbox{.}(2016){Reig}, {Nersesian}, {Zezas},
  {Gkouvelis}, \& {Coe}}]{reig16}
{Reig} P., {Nersesian} A., {Zezas} A., {Gkouvelis} L., {Coe} M.~J., 2016, \aap,
  590, A122

\bibitem[{{Rivinius}, {Carciofi} \& {Martayan}(2013){Rivinius}, {Carciofi}, \&
  {Martayan}}]{rivinius13}
{Rivinius} T., {Carciofi} A.~C., {Martayan} C., 2013, \aapr, 21, 69

\bibitem[{{Romoli} {et~al}\mbox{.}(2017){Romoli}, {Bordas}, {Mariaud},
  {Murach}, \& {H.~E.~S.~S. Collaboration}}]{romoli17}
{Romoli} C., {Bordas} P., {Mariaud} C., {Murach} T., {H.~E.~S.~S.
  Collaboration}, 2017, International Cosmic Ray Conference, 301, 675

\bibitem[{{Rouco Escorial} {et~al}\mbox{.}(2019){Rouco Escorial},
  {Hern{\'a}ndez Santisteban}, {Echevarr{\'\i}a}, {Wijnands}, {Page}, \&
  {Degenaar}}]{roucoescorial19}
{Rouco Escorial} A., {Hern{\'a}ndez Santisteban} J.~V., {Echevarr{\'\i}a} J.,
  {Wijnands} R., {Page} D., {Degenaar} N., 2019, Research Notes of the American
  Astronomical Society, 3, 31

\bibitem[{{Salas} {et~al}\mbox{.}(2013){Salas}, {Ma{\'\i}z Apell{\'a}niz},
  {Gamen}, {Barba}, {Sota}, {S{\'a}nchez-Berm{\'u}dez}, \& {Alfaro}}]{salas13}
{Salas} J., {Ma{\'\i}z Apell{\'a}niz} J., {Gamen} R.~C., {Barba} R.~H., {Sota}
  A., {S{\'a}nchez-Berm{\'u}dez} J., {Alfaro} E.~J., 2013, The Astronomer's
  Telegram, 5571, 1

\bibitem[{{Takata} {et~al}\mbox{.}(2012){Takata}, {Okazaki}, {Nagataki},
  {Naito}, {Kawachi}, {Lee}, {Mori}, {Hayasaki}, {Yamaguchi}, \&
  {Owocki}}]{takata12}
{Takata} J. {et~al.}, 2012, \apj, 750, 70

\bibitem[{{Takata} {et~al}\mbox{.}(2017){Takata}, {Tam}, {Ng}, {Li}, {Kong},
  {Hui}, \& {Cheng}}]{takata17}
{Takata} J., {Tam} P.~H.~T., {Ng} C.~W., {Li} K.~L., {Kong} A.~K.~H., {Hui}
  C.~Y., {Cheng} K.~S., 2017, \apj, 836, 241

\bibitem[{{Telting} {et~al}\mbox{.}(1994){Telting}, {Heemskerk}, {Henrichs}, \&
  {Savonije}}]{telting94}
{Telting} J.~H., {Heemskerk} M.~H.~M., {Henrichs} H.~F., {Savonije} G.~J.,
  1994, \aap, 288, 558

\bibitem[{{The Fermi-LAT collaboration}(2019)}]{lat_4fgl}
{The Fermi-LAT collaboration}, 2019, arXiv e-prints, arXiv:1902.10045

\bibitem[{{Zabalza}(2015)}]{naima}
{Zabalza} V., 2015, ArXiv e-prints

\bibitem[{{Zamanov} {et~al}\mbox{.}(2001){Zamanov}, {Reig}, {Mart{\'\i}},
  {Coe}, {Fabregat}, {Tomov}, \& {Valchev}}]{zamanov01}
{Zamanov} R.~K., {Reig} P., {Mart{\'\i}} J., {Coe} M.~J., {Fabregat} J.,
  {Tomov} N.~A., {Valchev} T., 2001, \aap, 367, 884

\end{thebibliography}

\label{lastpage}
\end{document}